\documentclass[journal,onecolumn]{IEEEtran}

\usepackage{amsmath,amsfonts}
\usepackage{algorithmic}
\usepackage{graphicx}
\usepackage{textcomp}
\usepackage{xcolor}
\usepackage{subcaption}
\usepackage[formats]{listings}
\usepackage{mdframed}
\begin{document}

\title{diaLogic: Non-Invasive Speaker-Focused Data Acquisition for Team Behavior Modeling}

\author{\IEEEauthorblockN{R. Duke}
\IEEEauthorblockA{\textit{Department of Electrical \& Computer Engineering} \\
\textit{Stony Brook University}\\
NY, USA \\
ryan.duke@stonybrook.edu}
\and
\IEEEauthorblockN{A. Doboli}
\IEEEauthorblockA{\textit{Department of Electrical \& Computer Engineering} \\
\textit{Stony Brook University}\\
NY, USA \\
alex.doboli@stonybrook.edu}
}

\maketitle

\begin{abstract}
This paper presents diaLogic system, a Human-In-A-Loop system for modeling the behavior of teams during solving open-ended problems. Team behavior is modeled through the hypotheses extracted from features computed from acquired voice data. These features include speaker interactions, speaker emotions, fundamental frequencies, and the corresponding text and clauses. Hypotheses about the invariant and differentiated situations are found based on the similarities and dissimilarities of the behavior of teams over time.  To provide full automation of data acquisition, the diaLogic system is executed within an intuitive, user-friendly GUI interface. Experiments present the performance of the system for a broad set of cases featuring team behavior during problem solving.
\end{abstract}


\section{Introduction}


Human-In-A-Loop (HiL) systems integrate data acquisition, modeling, and decision making for activities and process that tightly couple computing with human activities and characteristics, like emotions, social interactions, beliefs, goals, and so on~\cite{HiLS1}. The tight coupling of computing and humans impose intriguing new challenges in which the formalized, algorithmic, and well-defined nature of traditional computing must coordinate with human behavior, which is more spontaneous, less structures, and often open-ended. An important problem in this context is the {\em semantic coupling} between computing and human activities, where by semantic coupling we mean the capacity to establish a sound and comprehensive bi-directional way to communicate meaning between computers and humans. In particular, this paper discusses an algorithmic solution to modeling and representing human behavior during problem solving in teams~\cite{Doboli2021, Doboli2021b}.      

Data acquisition and modeling has been a traditional topic in Engineering. The challenge is to collect the maximum amount of information within the resources of the system and with minimal user input. 
Such systems can be implemented in two conceptual ways. Multi-input, single-output systems feature multiple data sources, which are organized into a single intuitive output data format, usually used for one purpose. Single-input, multi-output systems feature a single data source, which utilizes various algorithms to create multiple data outputs. These output data are usually used for multiple purposes, either simultaneously or selectively.
Within either process, models can be built at different points. For the top-down approach, parameterized models are created before data acquisition. Data is then used to find the model parameters. For the bottom-up approach, models are identified (i.e., selected, inferred, or synthesized) during data after acquisition. 
This work argues that the bottom-up method is the better approach for creating models for HiL systems.

This paper presents diaLogic system, a HiL system to model team behavior during open-ended problem solving. It is a single-input, multi-output automated data acquisition system that utilizes multi-faceted data to extract hypotheses about team behavior. Cognitive, emotional, and social aspects of teams are computed based on speech within recorded videos of social experiments, in which individuals interact with each other during problem solving. The addressed semantic coupling includes identifying the concepts and types of clauses forming the responses as well as the emotions and nature of team interactions.  

The model of the proposed system assumes that team behavior is a sequence of linear segments during which there are no significant changes of the following five sets: the set of concepts used in responses, the set of agent emotions, the set of observed urgency in creating responses, the set of observed motivations, and the differences between current and previous responses. The similarities and dissimilarities of linear segments are then used to extract hypotheses over different teams and time segments. Hypotheses correspond to two situations: (1)~invariant situations expressed by the similarities of the five sets pertaining to different linear segments and (2)~differentiated situations corresponding to the dissimilarities of the five sets. A rule-based algorithm identifies the types of the clauses in responses: {\em What}, {\em Who}, {\em For who}, {\em When}, {\em How}, {\em Where}, {\em Why}, and {\em Consequences} clauses.   diaLogic system is designed in Python with the PyQt5, Numpy, Keras, Azure, CoreNLP\cite{corenlp}, and PyWSD\cite{pywsd} libraries. The core algorithm for diaLogic is a standard speaker diarization algorithm, from which all subsequent data are generated, 
such as speech emotion recognition, speaker interaction, speech-to-text, and speech clause information. 

diaLogic system can be utilized in broad range of applications from education to team research and psychology. This system features higher accuracy when processing normal social conversation, rather than situations which feature conversations within a specialized context, like the jargon used in Computer Science. 

The paper has the following structure. Section~II discusses related work. Section~III presents diaLogic's hypotheses extraction algorithm. Section~IV details the system design, including speaker diarization, speaker interaction detection, speech emotion recognition, speech-to-text conversion, and speech clause detection. Experimental results are offered next. Conclusions end the paper.  

\section{Related Work}

This section discusses related work. Systems similar to diaLogic system are scarce, since most research focuses on the individual components of the system. Therefore, the related work section concentrates on the individual components.

{\em A. Speaker diarization}. Refinements to speaker diarization focus on improving the individual processing stages of the core algorithm. These stages include audio processing, neural network architecture, and spectral clustering. For audio processing, audio quality is optimized. Audio de-noising and audio de-reverberation have been utilized to remove background noise within a specific recording. These methods are implemented either through purely mathematical means or through machine learning~\cite{rp2-01} \cite{rp2-13}.

Neural network (NN) architecture refinements are some of the most common related work topics within speaker diarization. CNN networks, LSTM networks, or architecture combinations have been documented. Furthermore, specific layer-level improvements have been proposed. For example, optimizations to the softmax loss function are proposed to reduce error. Recurrent layers have been devised to reduce error as well \cite{rp2-12}. These optimizations are data-specific, and provide increased performance for only a specific subset of data. Therefore, as more optimizations are implemented, the more specialized the data becomes.

For the top-down approach to speaker diarization, refinements to spectral clustering have been widely documented. Cross-EM validation is the most common refinement. This feature improves the cross-validation process within the K-means clustering algorithm. Agglomerative Hierarchical Clustering (AHC) and Sequential Constraint (SC) clustering have been suggested as well. These refinements provide a specialized approach to minimizing clustering error \cite{rp2-09}.

{\em B. Interaction Characterization}. Related work for interaction characterization involves similar, but not directly related implementations. Interaction characterization through audiovisual means is the most common approach. This method does not characterize interactions through speech, but provides characterization through visual head and expression tracking~\cite{rp2-10}. This algorithm can be utilized to characterize sentiment analysis, where a person's visual expressions can merit approval or disapproval~\cite{rp2-08}.

Interaction characterization through audio data has been proposed, but it has not been automated in such an implementation. In one implementation, sentiment analysis had been recorded by hand from audio data~\cite{rp2-04}. In another implementation, speaker diarization data is automatically recorded, from which interaction data is then recorded by hand~\cite{rp2-07}. Recording interactions by hand is laborious and subjective to a person's rationale and opinions. A mathematical algorithm provides an accuracy reduction in exchange for faster processing, as well as a logical baseline for interaction characterization.

{\em C. Speech Emotion Recognition (SER)}. The determining factor for related work in SER is the database from which the NN for predictions is trained. Currently, non-proprietary databases do not yield accuracy above 93\% within training results. Therefore, the accuracy of related SER implementations is limited until new databases are created. The public SER database with the highest training accuracy is the German EMODB database~\cite{emodb}. This database features high quality samples, which contributes to its high accuracy. However, the extremely expressive nature of the German language also is a contributing factor to accuracy. Various implementations of EMODB have been designed. A Support Vector Machine (SVM) implementation achieved an 80\% accuracy~\cite{rp2-16}. A fused multi-class SVM implementation offers 93\% accuracy~\cite{rp2-17}. A CNN-based implementation achieved a 90\% accuracy~\cite{rp2-18}. An attention-based CNN \& LSTM implementation has 87\% accuracy~\cite{rp2-19}. A CNN-based implementation featured an 86\% accuracy \cite{rp2-20}. The accuracies from related implementations range from 80\% to 90\%. This indicates the importance of a high-quality SER database.

\begin{figure*}[t]
	\centering
	\includegraphics[width=120mm]{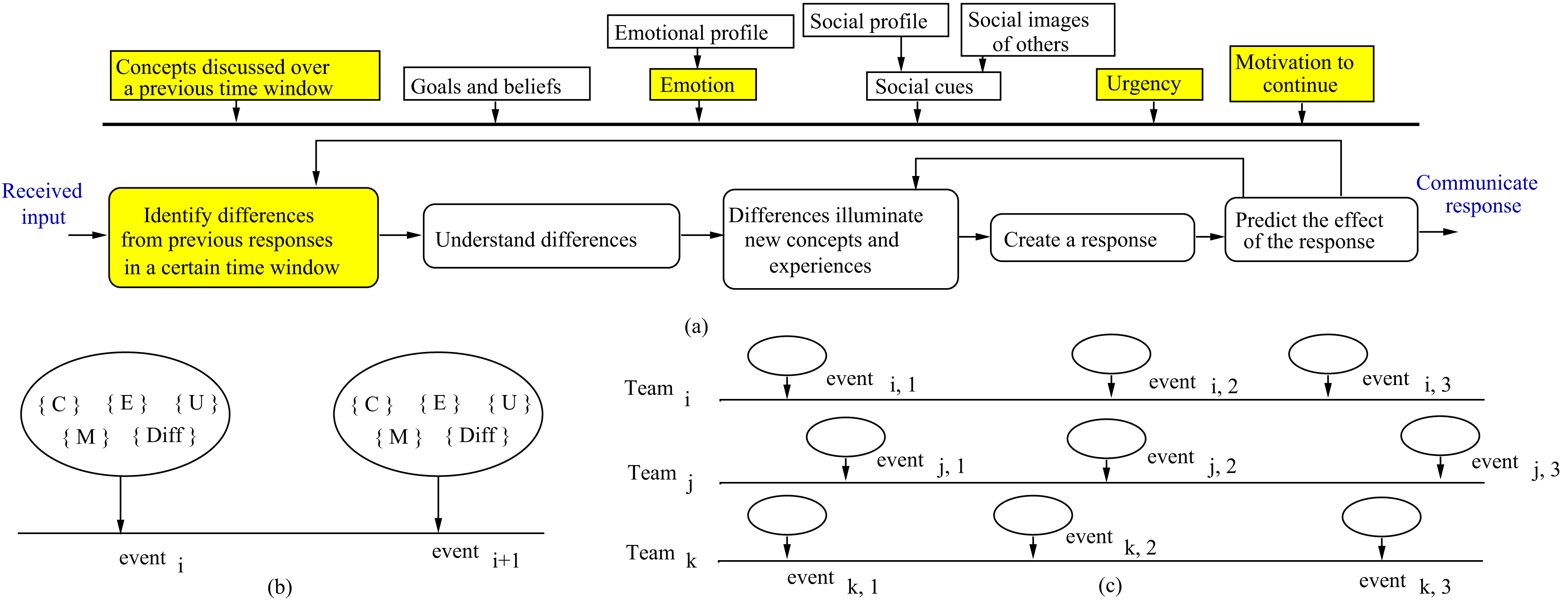}
	\caption{(a) Human agent model for HiL system, (b) event definition, and (c) team behavior comparison}
	\label{RD_1}
\end{figure*}

{\em E. Speech clauses detection}. The detection of speech clauses, or subordination clauses, within sentences has been studied too. The methods for determining these clauses fall into three categories: dependency parsing, grammatical role detection, and word-based analysis. These methods use a different approach to Natural Language Processing (NLP). 

For dependency parsing, the relationships between words within a sentence are utilized to generate hypotheses regarding sentence meaning. Algorithms which utilize dependency parsing seek to utilize knowledge beyond the parts of speech or individual word meanings, and determine the meaning of a sentence as a whole \cite{sc-03} \cite{sc-05}. Grammatical role detection algorithms utilize a simpler version of the former algorithm. The simplified approach seeks to extract specific parts of a sentence which fit into a specific grammatical context. Sentences which feature similar contexts are compared for further analysis. Furthermore, this algorithm can be combined with elements from the former algorithm to identify sentences which fit into a specific context, then extracts the parts of the sentences which are most prominent \cite{sc-01} \cite{sc-04} \cite{sc-06}. The most basic approach to speech clause detection focuses on a word-based approach. This approach focuses on determining the a sentence as a whole [22] [24], and is entirely dependent on the parts of speech assigned to each word. Speech clauses are detected by utilizing the lexical meanings of the words [21]. This final method features the highest dependency on context, and therefore the highest potential for error.

\begin{figure}[t]
	\centering
	\includegraphics[width=120mm]{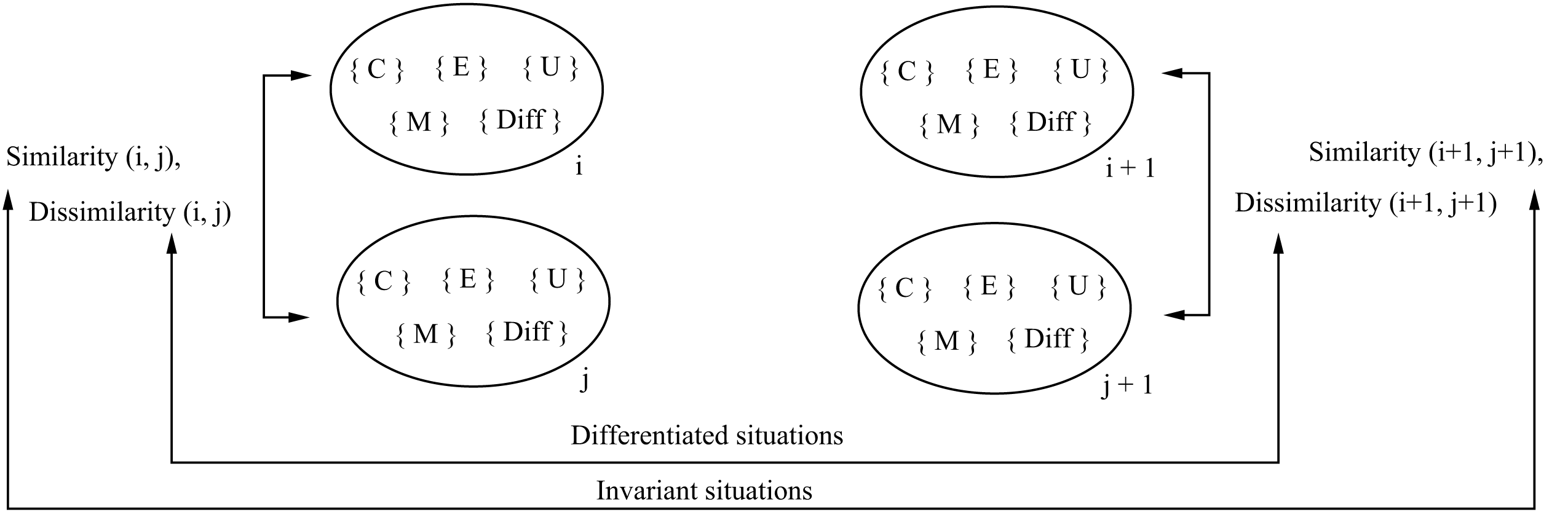}
	\caption{Hypothesis extraction principle}
	\label{RD_2}
	\vspace * {-0.1in}
\end{figure}

\section{Modeling Human-In-A-Loop Systems}

The data characterization and actuation of social systems is often referred to as a Human-Machine system or a Human-In-A-Loop (HiL) system~\cite{HiLS1}. The specification of these systems feature human input in the feedback process of the system design. HiL systems can directly interact with people through live data collection, or can be indirectly interactive through pre-recorded data. In either case, the system monitors individual subjects and characterizes data based on their actions. The data collected through these systems does not need to identify the names or IDs of the human subjects. Therefore, the data within these systems is usually anonymized. However, steps can be taken to create personalized data for each individual. This process increases the precision of the data collected, which allows the creation of personalized hypotheses.

Figure~\ref{RD_1}(a) summarizes the human agent model assumed for diaLogic HiL system~\cite{Doboli2021}\cite{Doboli2021b}. The bottom part of the model indicates the cognitive activities that an agent (i.e. participant) performs when communicating a response after receiving an input. It first identifies the differences between the current input and the previously received inputs and communicated outputs. Then, these differences are understood by the human agent, and as a result any related concepts and experiences are illuminated. These are used by the agent to create a response, but before communicating it, the agent predicts the expected effect of the response. Depending on the effect, the agent can decide to communicate the response to others, or, if unsatisfactory, new concepts and experiences might be illuminated or the agent might focus on other differences. 

The five cognitive activities are continuously moderated by the elements depicted in the upper part of Figure~\ref{RD_1}(a): the concepts discussed over the previous time window, the agent's goals and beliefs, and the agent's motivation to continue as a result of his/ her emotion, the interpretation of social cues, and the urgency (utility, valence) of the response. Not all components are observable. The figure highlights in yellow the activities and elements that can be directly tracked together with the responses communicated during team interaction. 

The observable behavior of a team of human agents is described as shown in Figure~\ref{RD_1}(b). At a given time moment, the team behavior is described by set $\{C\}$ of responses (e.g., concepts) mentioned over the last time window, set $\{E\}$ of emotions, set $\{U\}$ of observed urgency in creating responses, set $\{M\}$ of observed motivation, and set $\{Diff\}$ of differences between new and previous responses. The model defines that an {\em event} occurred if there is a significant difference between any of the five sets describing a team, e.g., at least one of the five sets at $event_i$ significantly differs from a set at $event_{i+1}$ in Figure~\ref{RD_1}(b). The time interval between two consecutive events is called {\em linear segment}.  

\begin{figure}[t]
	\centering
	\includegraphics[width=80mm]{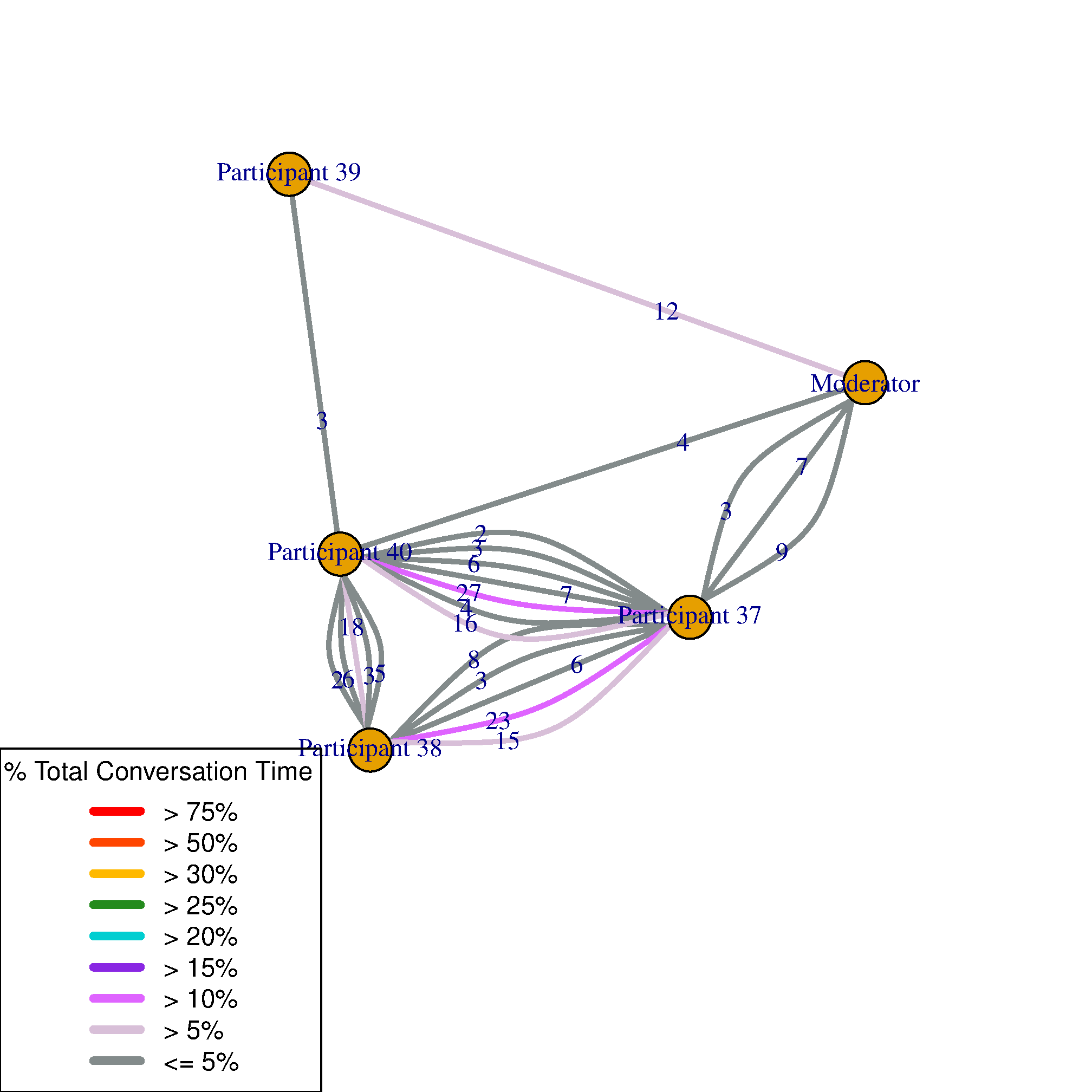}
	\caption{Interaction Graph}
	\label{I-1}
	\vspace * {-0.1in}
\end{figure}

\subsection {Hypotheses Extraction}

The types of hypotheses targeted in this work include invariant and consistent expressions that indicate how specific parameters affect the outcome of a team's behavior. Such invariant expressions refer to the individual sets $\{C\}$, $\{E\}$, $\{U\}$, $\{M\}$, $\{Diff\}$, relations among the parameters, and other parameters describing a team. {\em Invariance} is described by the property that under the same conditions, an expression holds over time and different teams. It supports finding the degree to which previous situations are a predictor to future cases for the same team or other similar team, e.g., Bayesian situations, as well as the degree to which a parameter change creates an expected behavior effect. Also, a new hypotheses must be {\em consistent} with hypotheses previously proved to be true, hence it should not logically contradict them.   

Figure~\ref{RD_2} summarizes the hypothesis extraction principle used in this work. It illustrates two linear segments, the first between the consecutive events $i$ and $i+1$ and the second between the consecutive events $j$ and $j+1$. Note that the two linear segments might pertain to the same team or to two different teams. Then, the methodology computes (using appropriate metrics) the similarity and dissimilarity between the five sets at the beginning and at the end of the linear segments as shown in the figure, e.g., between the sets for events $i$ and $j$ and for events $i+1$ and $j+1$. The extracted hypotheses represent two situations:
\begin{enumerate}
\item 
{\em Invariant situations}: They are described by the similarities of the pair 
$<~\{\{C\},$ $\{E\}, \{U\},$ $\{M\},$ $\{Diff\}\}_i,$ $\{\{C\},$ $\{E\},$ $\{U\}, \{M\},$ $\{Diff\}\}_{j}>$ and the similarities of the pair $<~\{\{C\}, \{E\},$ $\{U\}, \{M\},$ $\{Diff\}\}_{i+1},$ $\{\{C\}, \{E\},$ $\{U\}, \{M\}, \{Diff\}\}_{j+1}>$. It states that given the similarities between events $i$ and $j$, the similarities between events $i+1$ and $j+1$ were observed, in the presence of the dissimilarities between events $i$ and $j$ and the dissimilarities between events $i+1$ and $j+1$. As similarities exist in the presence of dissimilarities, the latter do not have an impact on the former. 

{\small
\begin{multline}
Sim(e_i, e_j) \implies \\ Sim(e_{i+1},e_{j+1}) | DSim (e_i, e_j), DSim(e_{i+1}, e_{j+1})   
\label{eq1}
\end{multline}
}
where $e_k$ indicates event $k$, $Sim$ is the similarity between two events, and $DSim$ is the dissimilarity. 
\item
{\em Differentiated situations}: They refer to the dissimilarities of the same pairs as invariant situations. However, they state that given the dissimilarities between events $i$ and $j$, the dissimilarities between events $i+1$ and $j+1$ were observed, in the presence of the similarities between events $i$ and $j$ and the similarities between events $i+1$ and $j+1$. Hence, the dissimilarities between events $i+1$ and $j+1$ result because of the dissimilarities between events $i$ and $j$, while the similarities might influence the magnitude of dissimilarities but not their existence.     
\end{enumerate}

{\small
\begin{multline}
DSim(e_i, e_j) \implies \\ DSim(e_{i+1},e_{j+1}) | Sim (e_i, e_j), Sim(e_{i+1}, e_{j+1})   
\label{eq2}
\end{multline}
}

\begin{figure}[t]
	\centering
	\includegraphics[width=70mm]{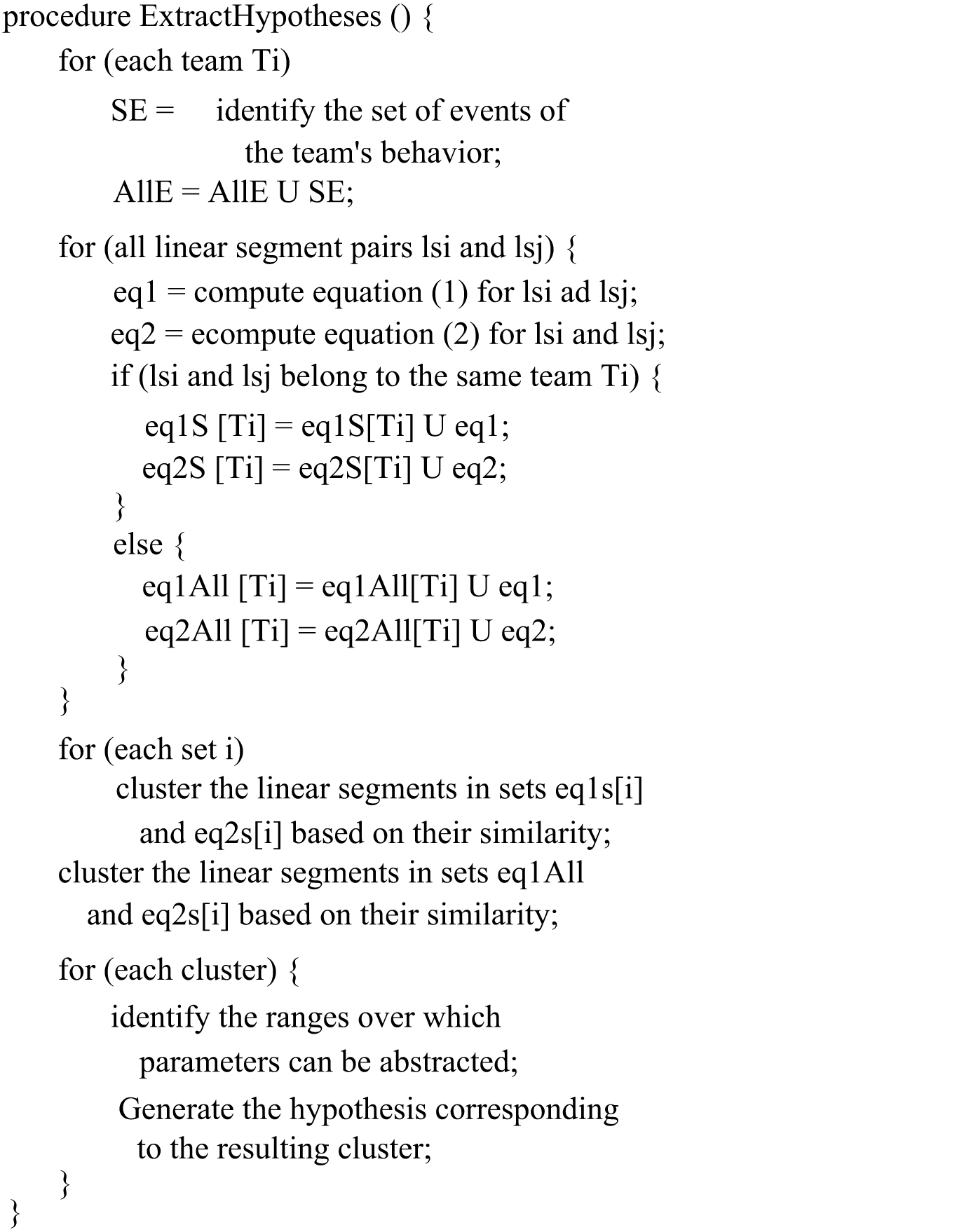}
	\caption{Hypothesis extraction algorithm}
	\label{RD_3}
	\vspace *{-0.1in}
\end{figure}

Figure~\ref{RD_3} presents the hypotheses extraction algorithm. The first step identifies the events in the behavior of each team $T_i$. Figure~\ref{RD_2}(c) illustrates the sequence of events, and the linear segments between two consecutive events for the same team. Set $AllE$ is the set of all linear segments in the considered data set.
For all pairs of linear segments $ls_i$ and $ls_j$ in set~$AllE$ equations~(\ref{eq1}) and~(\ref{eq2}) are computed and the results stored either in sets $eq1S$ and $eq2S$ corresponding to team $T_i$, if the linear segments pertain to the same team, or in the sets $eq1All$ and $eq2All$, if the linear segments are for different teams. Sets $eq1S$ and $eq2S$ are used to extract hypotheses for a certain team, while sets $eq1All$ and $eq2All$ are utilized to extract hypotheses for all teams. These sets are used as follows.

The maximum clustering of the linear segments are found for each set based on their similarity. Then, for each cluster, the ranges over which parameters can be abstracted are found by joining the values of the dissimilar parameters in equation~(\ref{eq1}). 

\begin{figure}[t]
	\centering
	\includegraphics[width=120mm]{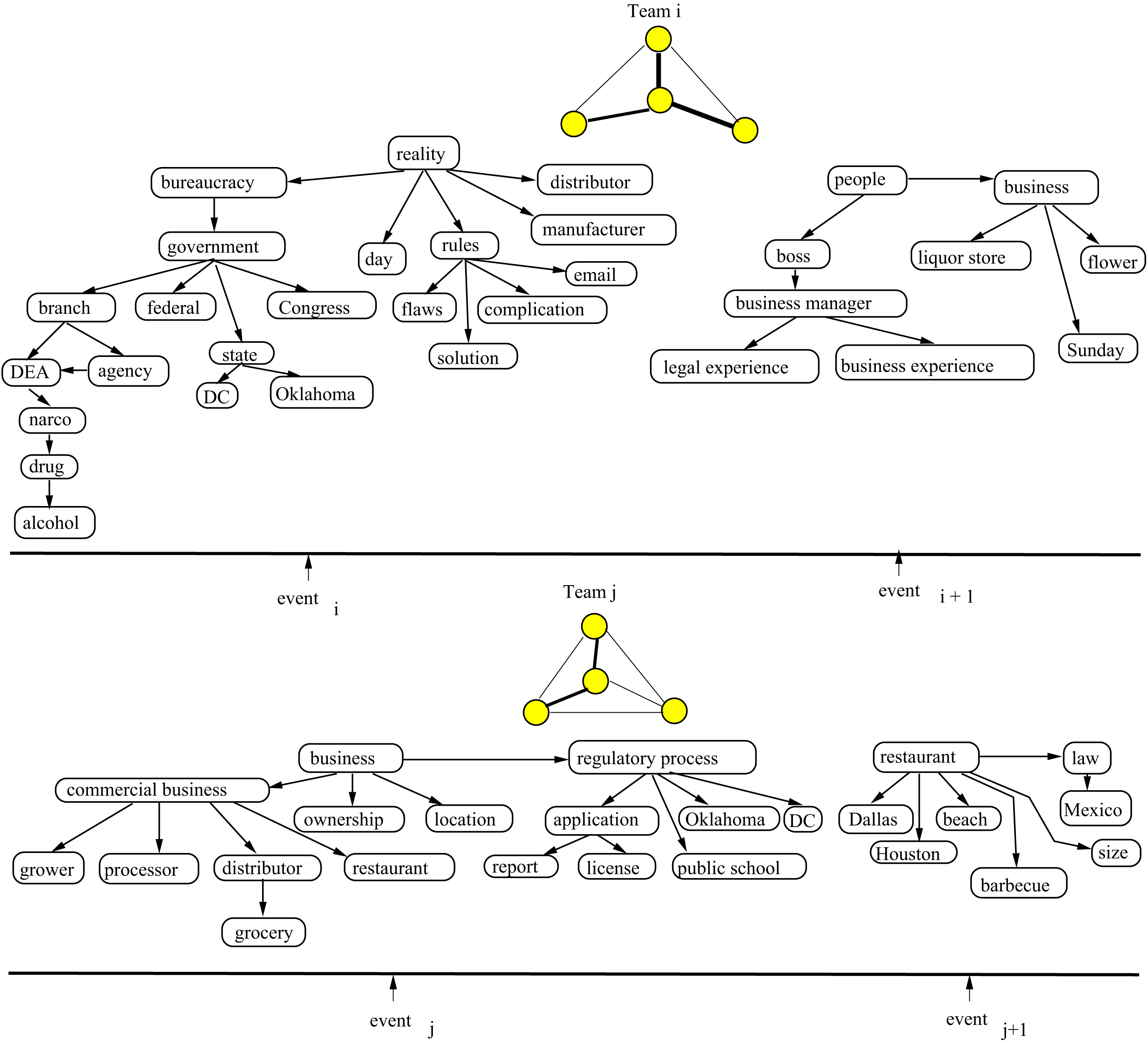}
	\caption{Example on hypothesis extraction}
	\label{RD_4}
	\vspace * {-0.1in}
\end{figure}

{\bf Example}: Figure~\ref{RD_4} illustrates hypothesis extraction for a case involving two teams that solve a problem on identifying a new business opportunity. The analysis explored the nature of the discussed ideas and their dependency on team characteristics, i.e. how does team interaction relate to the variety of the discussed concepts. Two consecutive events are shown for each team: events $i$ and $i+1$ for Team~$i$ and events $j$ and $j+1$ for Team~$j$. The figure presents the network of concepts~\cite{Doboli2015, Ferent2013} at every event, such as the concepts that were mentioned during the linear segment that ended at that event. For example, Team~$i$ referred to twenty two concepts during that linear segment, like bureaucracy, government, branch, federal, state, complications, and so on. The degree of relatedness between concepts is shown through the arcs connecting the concepts, like the arcs between concepts government and concepts federal, Congress, and state. The nature of the network indicates the breadth (diversity) of the concepts as well as the depth of the discussion, i.e. the degree of details. Equations~(\ref{eq1}) and~(\ref{eq2}) suggest that there is a similarity between events $i$ and $j$ as their concept networks shows similar breadths. However, the network for event~$j$ has a lesser depth as it includes only sixteen concepts. Events $i+1$ and $j+1$ are more similar both breath and depth-wise as both have nine and eight concepts, respectively. The interaction graphs show similar interaction patterns in the two teams: one member, shown as the central, yellow bubble, has strong interactions with all members, while the other members interact less with each other. 

The following equation~(\ref{eq1}) is extracted for this case:       
{\small
\begin{multline}
Sim(Br_{N_i}, Br_{N_j}) \implies Sim(Br_{N_{i+1}}, Br_{N_{j+1}}) \wedge \\ Sim(De_{N_{i+1}}, De_{N_{j+1}})\\ 
| Sim({IG_{i,i+1}, IG_{j,j+1}})   
\label{eq1_ex1}
\end{multline}
}

The hypothesis states that for concept networks with similar breadths ($Br$), similar breadths and depths ($De$) results if the two teams have also similar interactions ($IG$). 

The following equation (\ref{eq2}) is extracted for this example:

{\small
\begin{multline}
DSim(De_{N_i}, De_{N_j}) \implies  \bot \\ 
| Sim(Br_{N_i}, Br_{N_j}) \wedge Sim({IG_{i,i+1}, IG_{j,j+1}})   
\label{eq2_ex1}
\end{multline}
}

It states that the dissimilarity of the two concept network depths does not generate dissimilarities (denoted as $\bot$) at events $i+1$ and $j+1$, if the teams had initially the similar breadth of their concept networks and followed same interaction patterns. 

Note that hypotheses~(\ref{eq1_ex1}) and (\ref{eq2_ex1}) are invariants only for this example, therefore their validity for other situations needs to be verified. 




{\em Emotion and social interactions}. As explained in the next section, social interaction data can be integrated with emotion data to develop further hypotheses for emotional content. 
Changes in social interaction over time are indicated by changes in the Interaction Graphs $\Delta IG$, and changes in the emotional similarity between participants are $\Delta E$. Observing these changes allows extracting hypotheses about changes in group mood, changes in group interaction, and changes of individual's emotions. 

{\bf Example}: 
The data collected in an experiment showed that in a team of three participants, the two members with the least contribution to $\Delta IG$ (hence, the least social interaction) featured the least $\Delta E$. The third participant experienced the most $\Delta E$. The extracted hypothesis indicates that a participant that experienced a change of emotion also shows a change in interaction with the rest of the team.   

\subsection {Metrics, Similarity and Dissimilarity}

The following similarity and dissimilarity metrics are defined for each of the five sets $\{C\}$, $\{E\}$, $\{U\}$, $\{M\}$, and $\{Diff\}$. They are computed for individuals, such as the similarities and dissimilarities between two team members, or for entire teams, like the similarities and dissimilarities between two teams. 

{\em 1. Using concept networks (\{C\})}: The first set of metrics characterize concept networks~\cite{Doboli2015, Ferent2013}, like the number of concepts in the network, their variety (i.e. breadth of the network), the connections between concepts, the number of instances of more abstract concepts (e.g., the depth of the network), and the number of unrelated concepts in a network. Concept connections reflect the relationship between two concepts, such as using metrics obtained from WordNet~\cite{wordnet} or other similar databases. 

The second set of metrics considers the meaning (semantics) of the concepts in a response. The analysis determines the type of the clause in which a concept was used, like {\em What}, {\em How}, {\em Why}, {\em When}, {\em Who}, {\em For who}, and {\em Consequences} clauses. These metrics allow observing the nature of problem solving, e.g., the used reasoning. For example, metrics indicate if a certain participant or a team focused more on creating a solution (hence, there is a high number of {\em How} clauses) or focused more on problem framing (thus, used more {\em What} clauses). Section V.F details the rule-based algorithm that finds the type of the clauses.  

Finally, the third set of metrics refers to the intended meaning of a response, such as if the meaning corresponds to the meaning of the clauses, like using consequence clauses, or if the meaning is different due to the communicated intention, e.g., sarcasm, irony, confidence, or enthusiasm.

{\em 2. Using emotions (\{E\})}: These metrics characterize the degree to which emotions stay constant or change (transitions) and if certain emotions are predominant as compared to the others. Also, the similarity and dissimilarity of a member's emotions and those of the rest of the team are also computed. 

\begin{figure}[t]
	\centering
	\includegraphics[width=120mm]{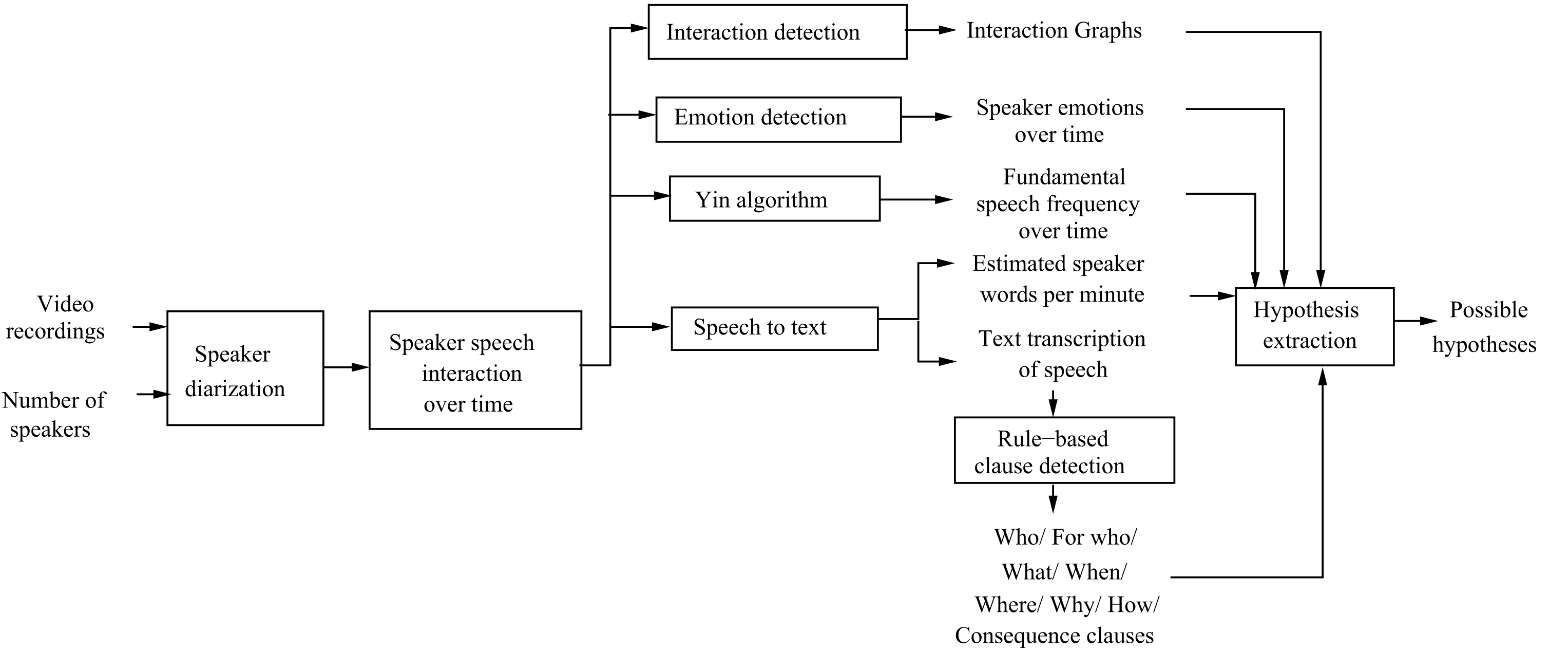}
	\caption{diaLogic system}
	\label{KG-1}
	\vspace * {-0.1in}
\end{figure}

{\em 3. Using urgency (\{U\})}: Metrics describe the priority of the produced responses. The priority reflects the outcomes expected by a team member, thus relate to his / her goals and beliefs. Urgency depends on the frequency of the responses an individual produces, the intention meant for the responses (like enthusiasm, confidence, sarcasm, etc.), and the emotions.  

\begin{figure*}[t]
	\centering
	\begin{subfigure}[h]{0.4\textwidth}
		\centering
		\includegraphics[width=\textwidth]{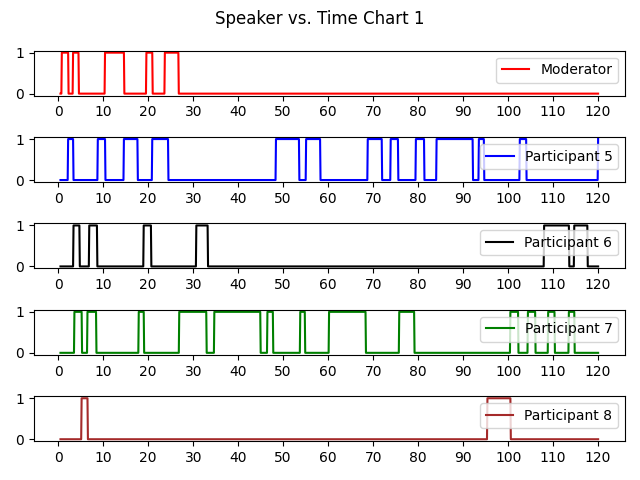}
		\caption{G2 T1}
		\label{ST1-1}
	\end{subfigure}
	\hfill
	\begin{subfigure}[h]{0.4\textwidth}
		\centering
		\includegraphics[width=\textwidth]{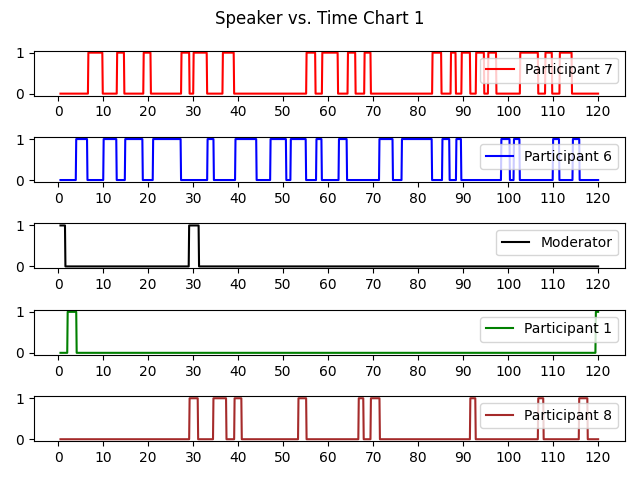}
		\caption{G2 T2}
		\label{ST1-2}
	\end{subfigure}
	\newline
	\begin{subfigure}[h]{0.4\textwidth}
		\centering
		\includegraphics[width=\textwidth]{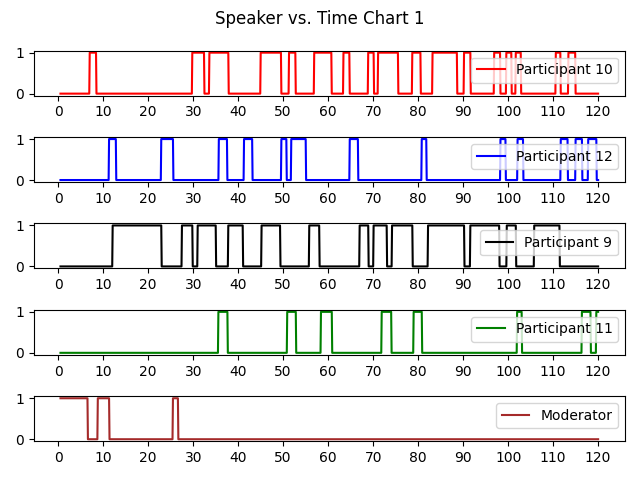}
		\caption{G3 T1}
		\label{ST1-3}
	\end{subfigure}
	\hfill
	\begin{subfigure}[h]{0.4\textwidth}
	\centering
	\includegraphics[width=\textwidth]{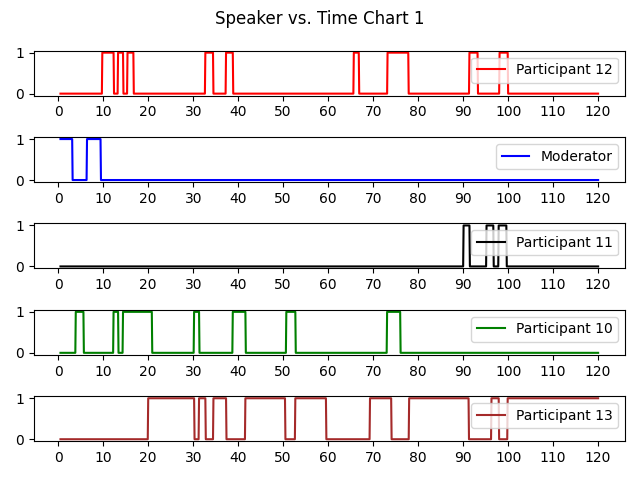}
	\caption{G3 T2}
	\label{ST1-4}
	\end{subfigure}
	\caption{First two minutes of speech from group videos G2-G3}
	\label{ST1}
\end{figure*}

{\em 4. Using motivations (\{M\})}: These metrics complement the metrics on urgency to express the degree to which a participant decides to allocate resources (i.e. attention, time, and energy) to address the perceived priority of a response. Motivation is characterized by the length of the responses, the nature of the used clauses, the variety of the referred concepts, and the emotions. Note that motivation correlates to the goals and beliefs of an individual.   

{\em 5. Using differences (\{Diff\})}: Metrics present the degree to which the metrics of the four previous types remained the same or changed. The following kinds of metrics are computed for the following five measurements: (i)~direct measurements are metrics extracted during data collection, (ii)~changes of direct measurements over time, (iii)~statistical correlations between direct measurements and changes, (iv)~sequences of situations that produce a certain outcome, and (v)~situations in which the computed metrics are insufficient to explain an outcome.   

\section {System Design}

This section presents the design and implementation of diaLogic system. 
Figure \ref{KG-1} illustrates the system. The inputs for this system are the video or audio file recordings of group interactions, with the number of speakers in each recording as a marker. Speaker diarization is performed on the audio data. The resulting data from speaker diarization drives every other processing steps, which include speech emotion detection, speaker interaction detection, speech-to-text conversion, words-per-minute computation, and speech clause detection. The data from the secondary datasets are used to generate hypotheses regarding team behavior.

\subsection{Speaker diarization}

\begin{figure}[t]
	\centering
	\begin{subfigure}[h]{60mm}
		\centering
		\includegraphics[width=40mm]{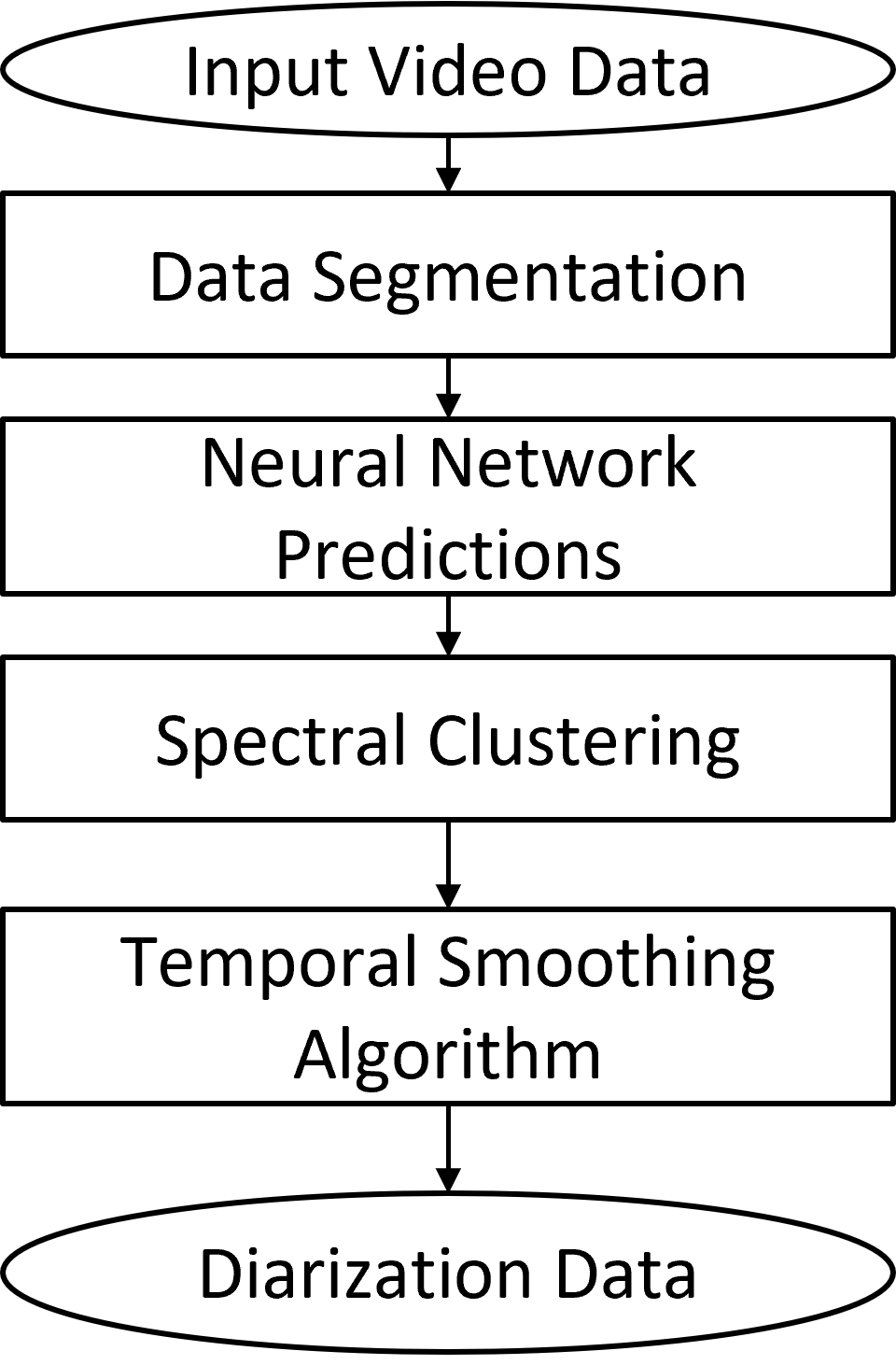}
		\caption{Speaker diarization procedure}
		\label{SD-F}
	\end{subfigure}
	\hfill
	\begin{subfigure}[h]{60mm}
		\centering
		\includegraphics[width=60mm]{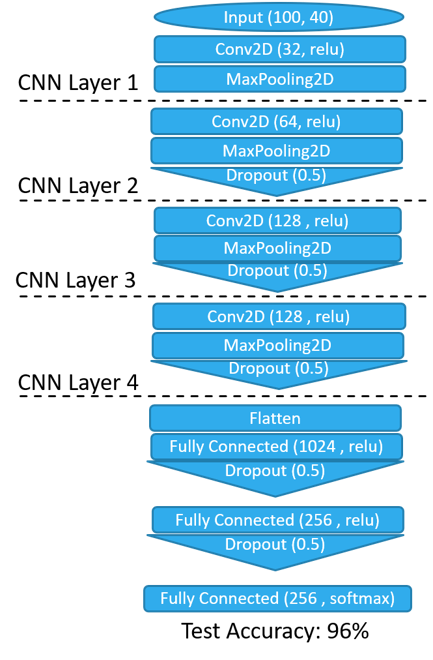}
		\caption{Speaker diarization CNN layout}
		\label{SD-CNN}
	\end{subfigure}
	\caption{Speaker diarization algorithm}
	\label{SD}
\end{figure}

Speaker diarization determines ``who spoke when'' within a given audio or video file. Figure~\ref{ST1} illustrates the output of diarization. Since this algorithm is the base for the other algorithms, the goal was to create an algorithm which functions adequately for future data interpretation. Hence, a relatively standard top-down procedure was designed based on four steps: data segmentation, neural network predictions, spectral clustering, and temporal smoothing. Figure \ref{SD-F} shows this process.

\begin{figure}[t]
	\begin{mdframed}
		\caption{Temporal Smoothing Algorithm}
		\label{pseudo}
		\begin{lstlisting}[language=C, basicstyle=\tiny]
Given: any speaker would speak consecutively for a minimum of 1.0s.
Therefore, for every speaker label: 
if a spike in speaker activity less than 1.0s in duration is followed 
by any further speech less than 1.0s later, then the two embeddings are merged. 
Else, if any spike in speaker activity less than 1.0s is followed by a 1.0s 
gap in speech, then the spike is negated. 
		\end{lstlisting}
	\end{mdframed}
\end{figure}

\begin{figure*}
	\centering
	\begin{subfigure}[h]{0.4\textwidth}
		\centering
		\includegraphics[width=\textwidth]{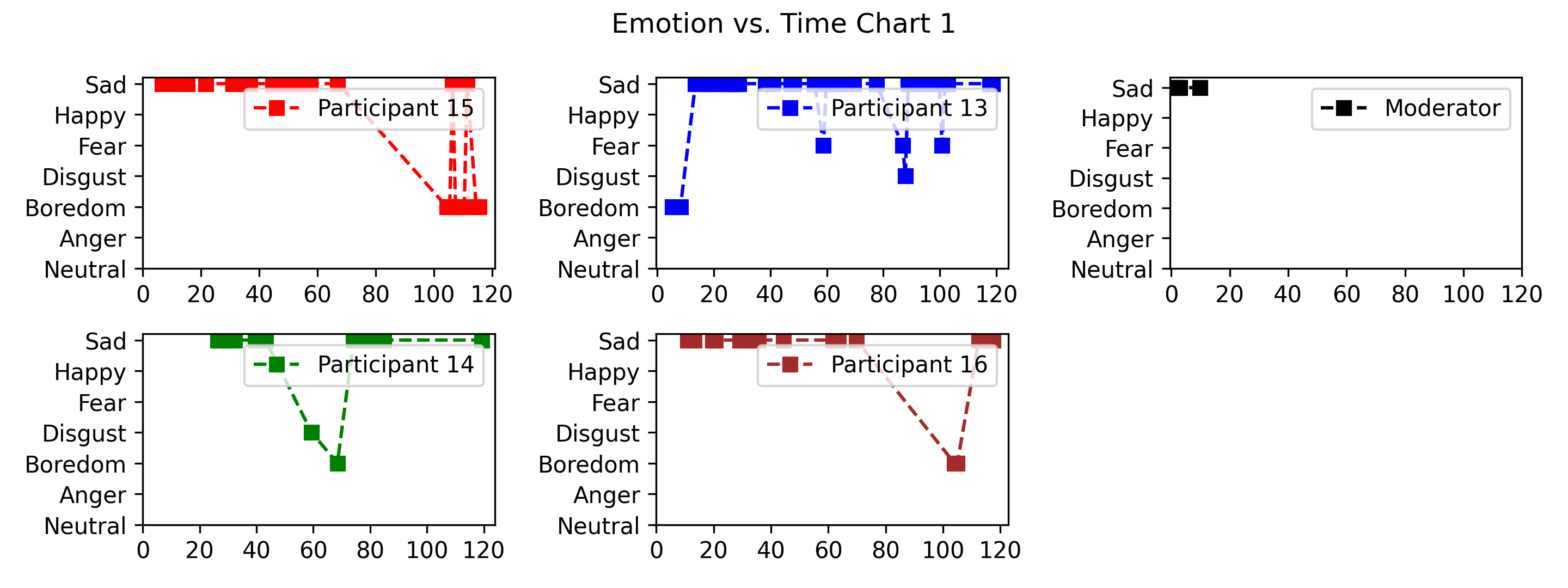}
		\caption{\textit{T}: 0-2 minutes}
		\label{SE1-1}
	\end{subfigure}
	\hfill
	\begin{subfigure}[h]{0.4\textwidth}
		\centering
		\includegraphics[width=\textwidth]{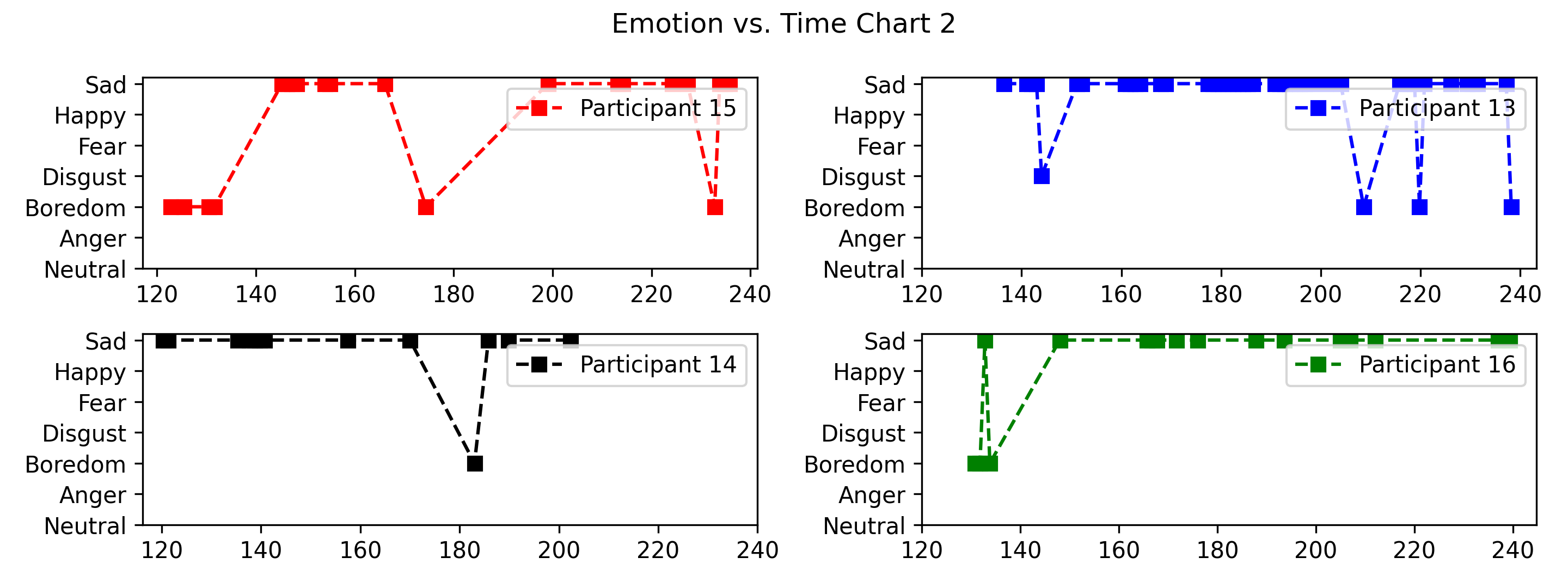}
		\caption{\textit{T}: 2-4 minutes}
		\label{SE1-2}
	\end{subfigure}
	\newline
	\begin{subfigure}[h]{0.4\textwidth}
		\centering
		\includegraphics[width=\textwidth]{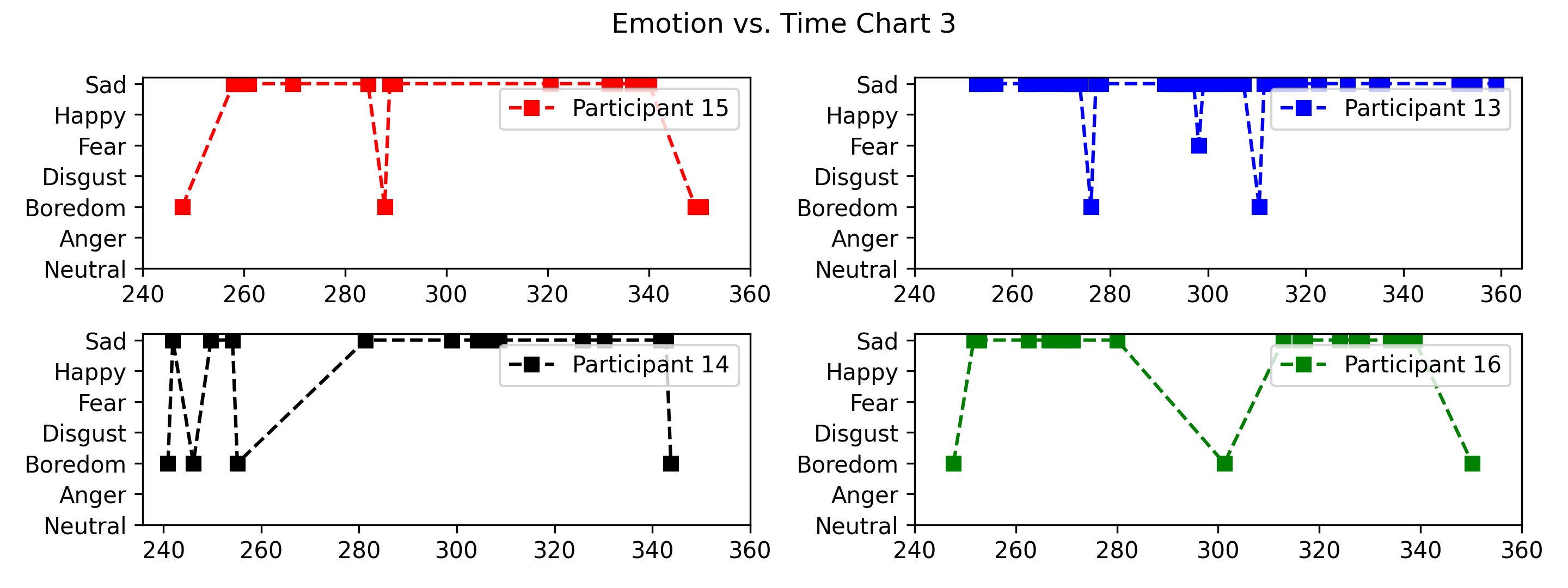}
		\caption{\textit{T}: 4-6 minutes}
		\label{SE1-3}
	\end{subfigure}
	\hfill
	\begin{subfigure}[h]{0.4\textwidth}
		\centering
		\includegraphics[width=\textwidth]{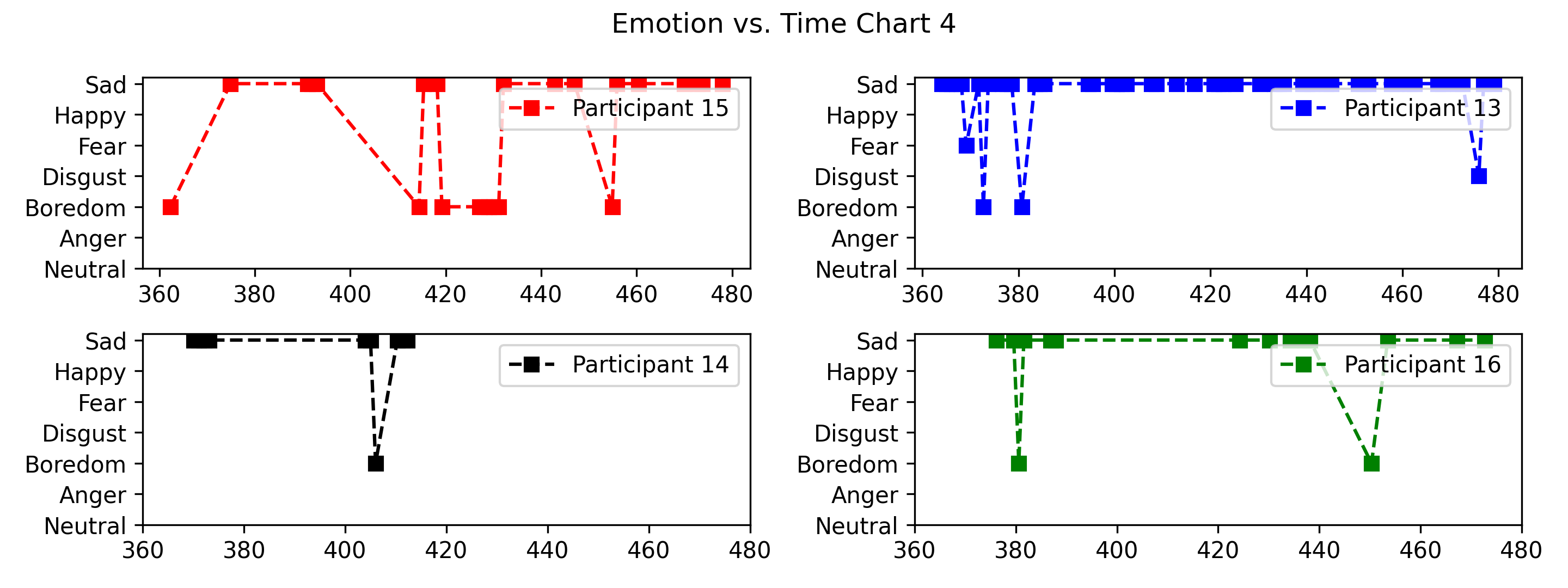}
		\caption{\textit{T}: 6-8 minutes}
		\label{SE1-4}
	\end{subfigure}
	\caption{G4 T1 emotion trends}
	\label{SE1}
\end{figure*}

\begin{figure}[t]
	\centering
	\begin{subfigure}[h]{60mm}
		\centering
		\includegraphics[width=40mm]{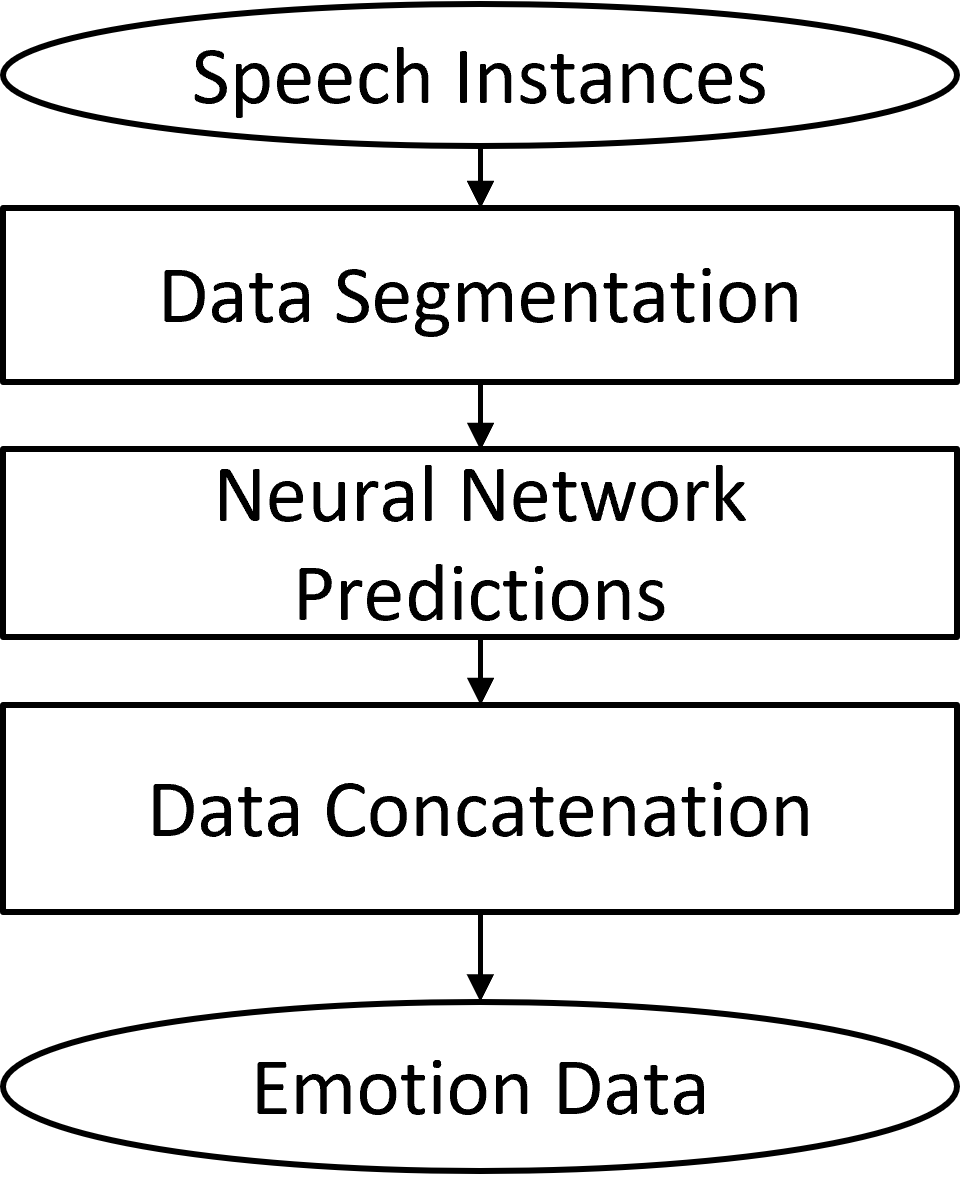}
		\caption{Speech Emotion Recognition procedure}
		\label{SER-F}
	\end{subfigure}
	\hfill
	\begin{subfigure}[h]{60mm}
		\centering
		\includegraphics[width=60mm]{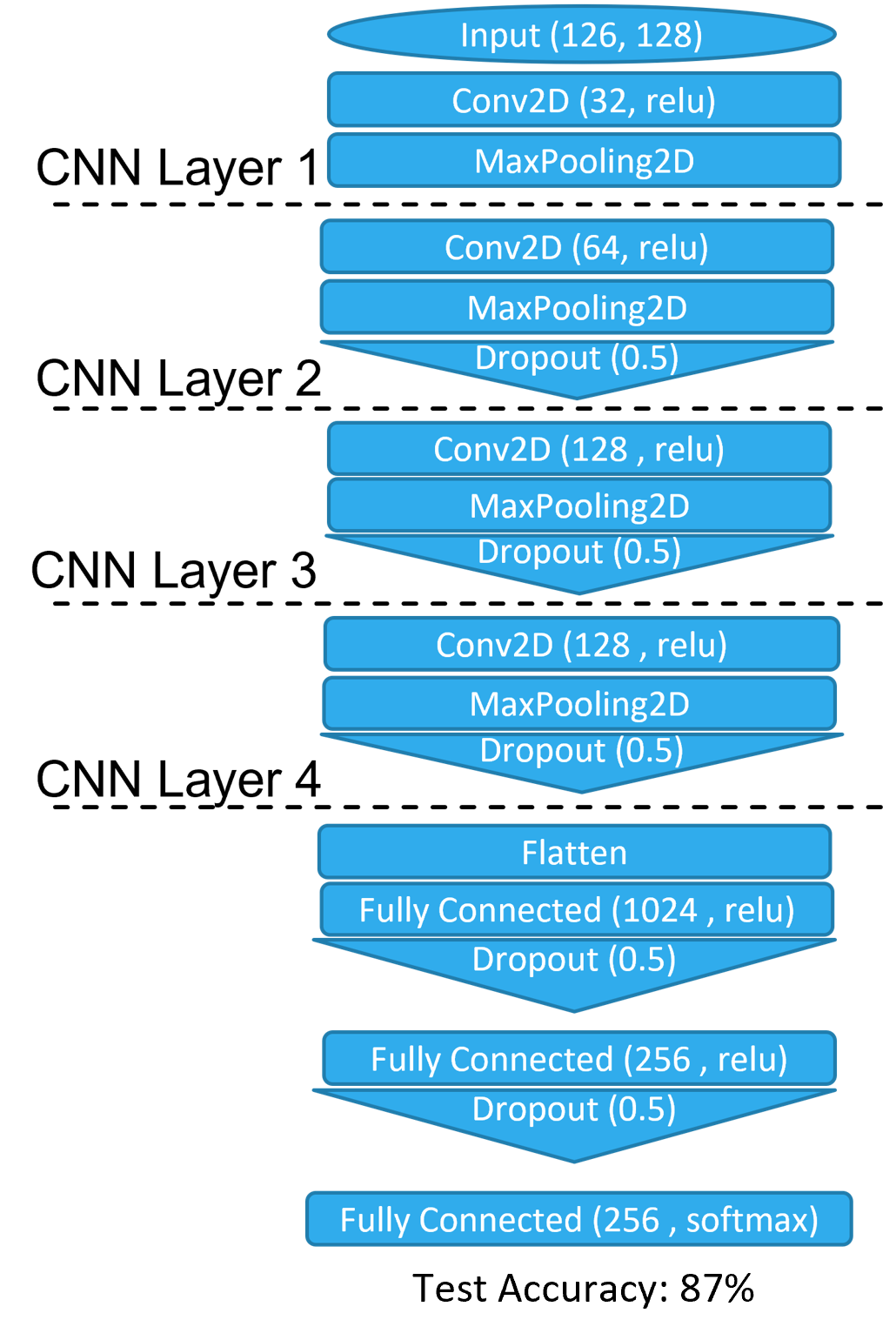}
		\caption{Speech Emotion Recognition CNN layout}
		\label{SER-CNN}
	\end{subfigure}
	\caption{Speaker Emotion Recognition (SER) algorithm}
	\label{SER}
\end{figure}

Data segmentation utilizes mel-scaled spectrograms for our data format, and audio files are segmented into 1~second utterances overlapping by 0.1~second. Each resulting mel spectrogram has dimensions of $(100, 40)$. Each spectrogram is input to a 4-layer CNN designed in the Python Keras library. The architecture of this network is shown in Figure \ref{SD-CNN}. This CNN was trained on 30 random speakers from the Librispeech audio corpus \cite{librispeech} with a final test accuracy of 96\%. Each prediction D-Vector is merged into an affinity matrix, and input to a spectral offline clustering algorithm, which was derived from the material in \cite{lstm}. The spectral clustering algorithm calculates the eigenvalue and eigenvectors of the affinity matrix, and clusters them into $m$ clusters using K-means, where $m$ was the number of speakers. 
The resulting data from spectral clustering is a series of utterances for each speaker, which denote a start and end time. 

In context-specific circumstances, utterances are personalized to match speaker identities. In every video set, each speaker introduces themselves sequentially at the start of the video. A separate text file is input for each video to list the speaker's IDs in order of introduction. Then, the IDs to each of the utterances' are assigned random IDs chronologically after spectral clustering was performed. 

The utterances generated from spectral clustering are unrefined, and can contain small spikes and overlaps in the speech of individual speakers. Therefore, a temporal smoothing algorithm removes spikes in speech. The logic for the temporal smoothing algorithm is presented in Figure \ref{pseudo}. This algorithm balances speaker utterances and provides a more accurate representation of speech over time. This data is the final data output of the speaker diarization algorithm. From this data, speaker vs. time charts can be created. One speaker vs. time chart is generated for every two minutes of audio within the list of utterances.

\subsection{Speaker interaction detection}

The utterances generated by speaker diarization are represented by an $N$ x $3$ array, where $N$ is the number of utterances and $3$ designates the utterance components: the speaker ID, the start time, and the end time. For every pair of speech instances within the $N$ x $3$ array, the conversation length is computed as the difference between the second end time and the first start time. The speaker is designated as the first ID, and the receiver as the second ID. Each of these interactions is recorded and documented in an Interaction Graph (IG). 
An IG represents every speaker as a node, every interaction as an edge, and the interaction time in seconds as a weight. Figure~\ref{I-1} illustrates an IG. One IG is generated for every matching two minute interval within each speaker vs. time chart. A final IG is produced for the entire video. 

This algorithm is a circumstantial interaction algorithm. It operates under the assumption that consecutive speech segments indicate an interaction between two people. It does not characterize intention, in the instance that a participant intends their message to be received by another. Instances where participants interrupt each other is a common occurrence of where this algorithm falls short.

\subsection{Speech Emotion Recognition}

The general objective of Speech Emotion Recognition (SER) is to determine the emotions of a given speaker within a given audio segment. A traditional SER algorithm was modified to integrate emotion data with the speaker-specific data from speaker diarization. The modified algorithm must be compatible with speaker diarization by featuring the same 1-second utterances over time, and offer output data that corresponds to the same 2 minute time intervals. The logical overview for SER is shown in Figure~\ref{SER-F}. The $N$ x $3$ array corresponding to speaker utterances is the input to the algorithm. Sequential data segmentation is performed on the individual utterances. This segmentation method is different from the method used in speaker diarization, as the segmented utterances do not overlap. However, the algorithm still utilizes mel-spectrograms. The spectrograms for SER are expanded and optimized for frequency response with a final dimension of $(126, 128)$. The segmented audio data is then used as prediction inputs to a secondary 4-layer CNN.

The CNN used for SER was trained on 783 random utterances from seven emotions within the EMODB dataset: \textit{Neutral, Anger, Boredom, Disgust, Fear, Happy, and Sad}\cite{emodb}. This CNN yielded a test accuracy of 87\%. The architecture of the CNN is displayed in Figure~\ref{SER-CNN}. When used for predictions, one emotion is derived for every consecutive second of speech. The emotion data are combined with the time-based data from each of its corresponding speech segments and concatenated. The resulting emotion data over time is used to plot emotion vs. time charts for every 2 minute interval of speech. The SER algorithm is simpler than its speaker diarization counterpart, as neither spectral clustering nor smoothing are needed. However, when the data generated are combined with the interaction data, further emotional interpretation hypotheses can be drawn.

\subsection{Speech-to-text conversion}

The speech-to-text conversion module of diaLogic system strives to add Natural Language Processing (NLP) components to the output data. Each detected utterance from speaker diarization is input to an online speech-to-text module to create speaker-specific text transcriptions. Microsoft Azure~\cite{azure} speech-to-text library is utilized due to its superior accuracy over the Google~API~\cite{google}. Since the privacy and identity of the speakers must be preserved, the first 30 seconds of each video are removed to eliminate the part in which speakers identified themselves, and then the utterances are taken out before processing with the Azure API. Both steps ensure that from the cloud standpoint, the utterances cannot be reassembled into a proper conversation.

The output of this module is an $N_s$ x $2$ array, where $N_s$ is the number of utterances minus the first 30 seconds, and value 2 designates the data, which is the speaker ID and the text transcription. For each utterance, the original speech duration can be utilized along with the number of words to estimate the words-per-minute rate for each utterance. Each individual rate is then averaged to estimate the average speech rate for each speaker across the entire video. Each of these metrics is output as a CSV file for future analysis.

\subsection{Speech Clauses Detection}

diaLogic system's goal in NLP is to detect the linguistic characteristics of sentences within speech. Within a sentence, the system consideres that specific linguistic properties represent subjects, and that the combination of those subjects indicates intention. The subjects consist of \textit{Who, For Who, What, When, Where, How,} and the intention as \textit{Why} and \textit{Consequences}. Speech-clause detection uses a rule-based algorithm based on CoreNLP and pyWSD libraries \cite{corenlp} \cite{pywsd}. These libraries are an extension of WordNet dictionary database \cite{wordnet}.

\begin{figure*}[t]
	\centering
	\begin{subfigure}[h]{0.4\textwidth}
		\centering
		\includegraphics[width=\textwidth]{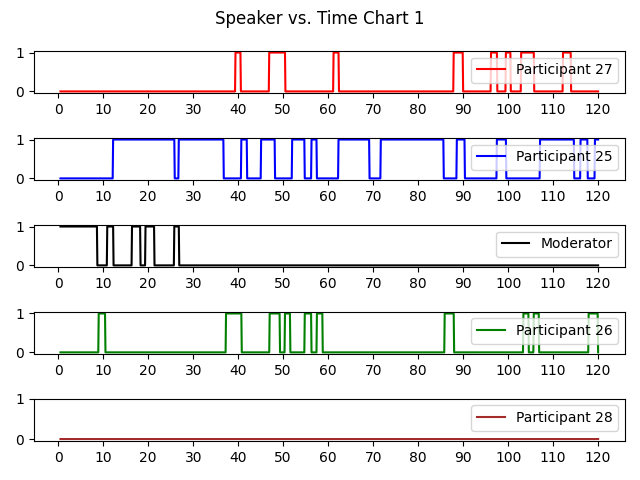}
		\caption{\textit{T}: 0-2 minutes}
		\label{ST2-1}
	\end{subfigure}
	\hfill
	\begin{subfigure}[h]{0.4\textwidth}
		\centering
		\includegraphics[width=\textwidth]{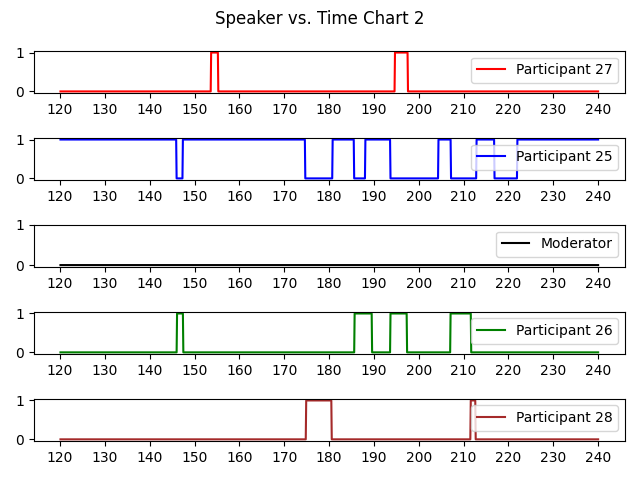}
		\caption{\textit{T}: 2-4 minutes}
		\label{ST2-2}
	\end{subfigure}
	\newline
	\begin{subfigure}[h]{0.4\textwidth}
		\centering
		\includegraphics[width=\textwidth]{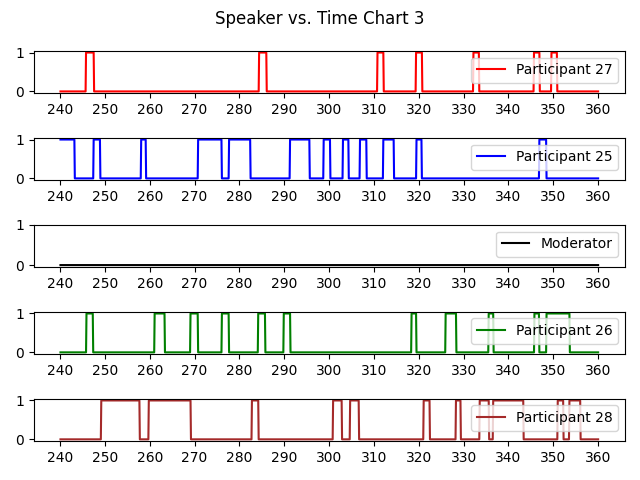}
		\caption{\textit{T}: 4-6 minutes}
		\label{ST2-3}
	\end{subfigure}
	\hfill
	\begin{subfigure}[h]{0.4\textwidth}
		\centering
		\includegraphics[width=\textwidth]{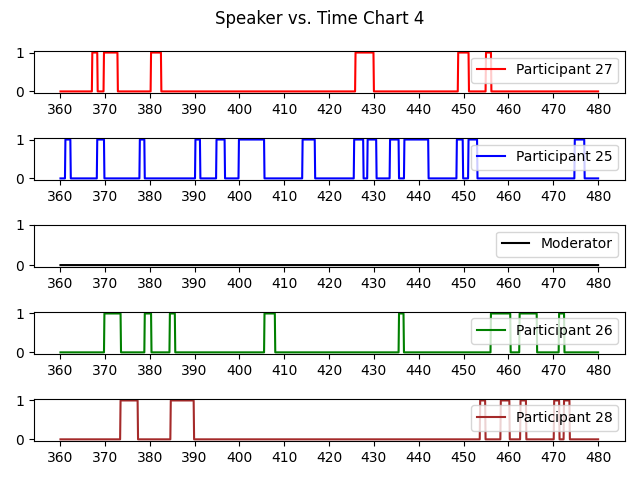}
		\caption{\textit{T}: 6-8 minutes}
		\label{ST2-4}
	\end{subfigure}
	\caption{G7 T1 speech trends}
	\label{ST2}
\end{figure*}

The algorithm finds the clauses in a sentence using the part of speech of each word in the sentence. There are a few requirements to this approach. First, the algorithm must accurately handle word ambiguity within a context. Second, it must be able to handle outlier cases, where words or sentences cannot be processed. Finally, the data must be represented in an intuitive format. To resolve ambiguity, CoreNLP1~\cite{corenlp} finds the parts of speech, and pyWSD~\cite{corenlp} determines the category of a word. Different parts of speech represent different functions within the algorithm. Verbs are considered to be actions and marked as anchor points. Nouns are disambiguated using pyWSD to find the meaning of the word within the sentence context, and then to assign it to a category, which represents a WordNet synset~\cite{wordnet}. Each category corresponds to each of the detected subjects: clauses~{\em Who} and~{\em For Who} correspond to the \textit{PERSON} category, clause~{\em What} is represented by \textit{ORGANIZATION and MISC} categories, clause {\em When} relates to the \textit{DATE, TIME, DURATION}, or \textit{SET} categories, and clause {\em Where} is represented by the \textit{LOCATION} category. Clause {\em How} is represented by a word that is either an adjective or an adverb. 

The steps (rules) of the algorithm are as follows:
(i)~First, the algorithm breaks down each transcription at the sentence level. (ii)~For every verb in a sentence, it looks at the words around them. (iii)~If a noun with the \textit{PERSON} label comes before the first verb in a sentence, the word is marked as a clause {\em Who}. (iv)~If the same noun comes after the first verb, the word is labeled as a clause {\em For Who}. (v)~The first instance of words in the clauses {\em What, Where, When} are recorded. (vi)~Descriptor words for clause~{\em How} are recorded if they come after a verb in a sentence, and one instance of descriptor words is recorded for each verb. Each of these words are double-marked with the verb that proceeds them, and the verb that follows them. This word-tagging method is used to determine clauses~{\em Why} and {\em Consequences}.

Finding clauses {\em Why} and {\em Consequences} considers the natural structure of sentences. (vii)~Clauses {\em Why} were represented as the following sentence: {\tt Because [blank] [verb] [blank] [descriptor]}. Some examples of a coherent {\em Why} sentence is Because \textbf{they} did \textbf{programming well}, or Because \textbf{I} took \textbf{courses sparingly}. One clause {\em Why} is built for each verb in the original sentence. The first {\tt blank} is filled by the noun recorded within the first five speech clauses that is before the specified verb. The second {\tt blank} is filled by the noun recorded within the first five speech clauses that is after the specified verb. {\tt descriptor} is filled by clause {\em How} which occurs after the verb. (viii)~For clause {\em Consequences}, the verb along with the noun in the second {\tt blank} are formed into a separate sentence respectively. From the examples above, the consequences would be \textbf{did programming}, and \textbf{took courses}. One clause {\em Consequences} is built for every verb in the original sentence.

\begin{figure*}[t]
	\centering
	\begin{subfigure}[h]{0.4\textwidth}
		\centering
		\includegraphics[width=\textwidth]{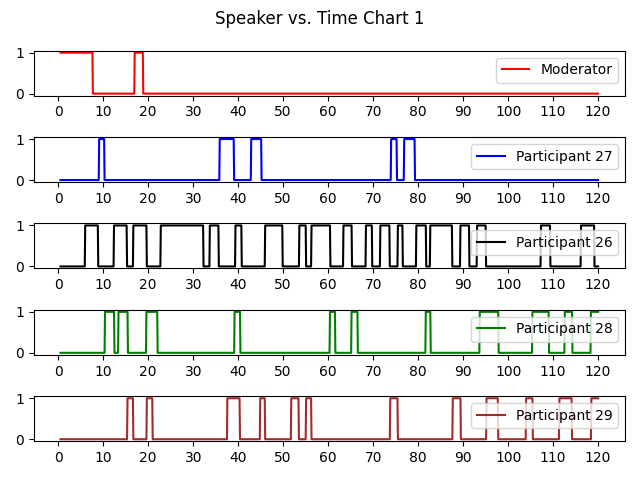}
		\caption{\textit{T}: 0-2 minutes}
		\label{ST3-1}
	\end{subfigure}
	\hfill
	\begin{subfigure}[h]{0.4\textwidth}
		\centering
		\includegraphics[width=\textwidth]{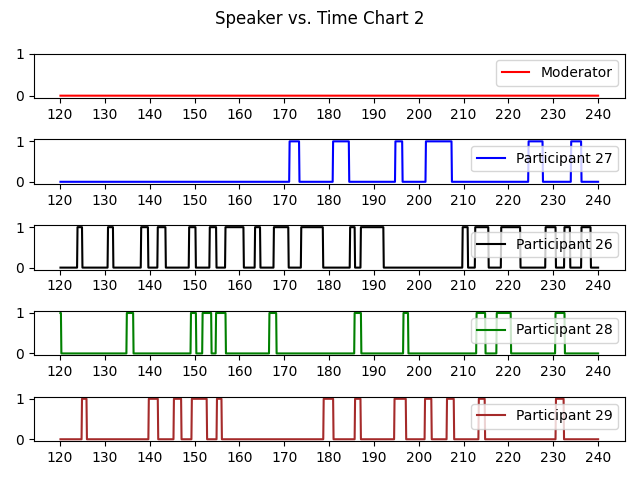}
		\caption{\textit{T}: 2-4 minutes}
		\label{ST3-2}
	\end{subfigure}
	\newline
	\begin{subfigure}[h]{0.4\textwidth}
		\centering
		\includegraphics[width=\textwidth]{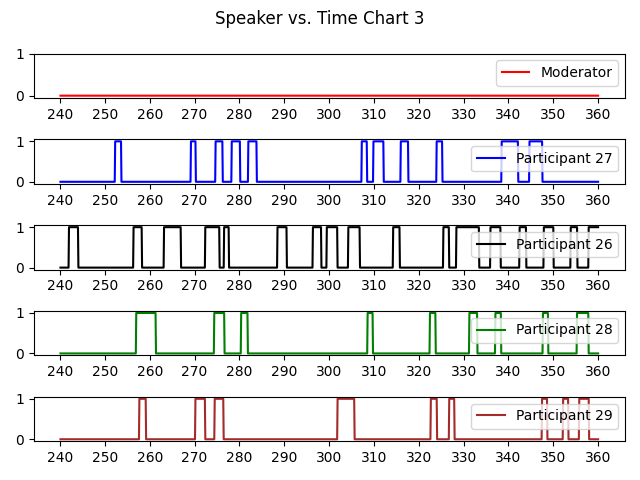}
		\caption{\textit{T}: 4-6 minutes}
		\label{ST3-3}
	\end{subfigure}
	\hfill
	\begin{subfigure}[h]{0.4\textwidth}
		\centering
		\includegraphics[width=\textwidth]{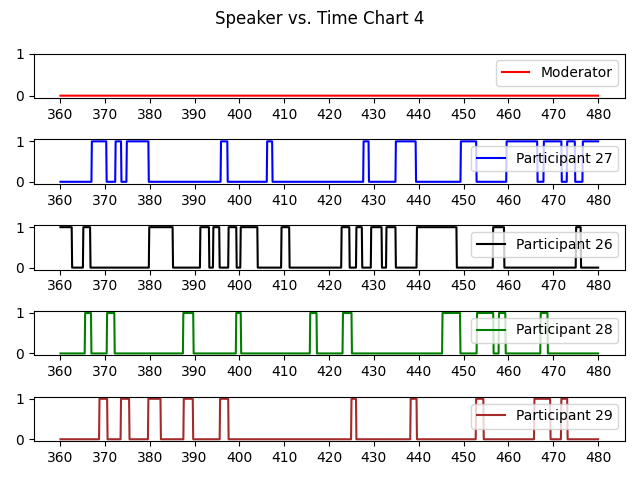}
		\caption{\textit{T}: 6-8 minutes}
		\label{ST3-4}
	\end{subfigure}
	\caption{G7 T2 speech trends}
	\label{ST3}
\end{figure*}

Due to the context of each sentence, the nature of the detected speech clauses can vary. Not all speech clauses will be detected in every sentence, and consequently, clauses {\em How}, {\em Why}, and {\em Consequences} will not be generated for every verb. Furthermore, some sentence may not contain verbs. Due to these circumstances, logic must be produced to handle outlier cases. The program treats every utterance as a single entity, but multiple sentences can exist within a single utterance. (ix)~For a sentence which does not contain verbs in a multi-sentence utterance, the sentence will be added to either the preceding or following sentence, and the speech clauses will be detected for the combined sentence. (x)~For sentences without verbs in single-sentence utterances, the sentence is ignored. (xi)~For utterances which do not contain text, the utterance is ignored.





\begin{table}[t]
\caption{Interruption frequency analysis of interaction characterization}
\begin{center}
\begin{tabular}{|c|c|c|c|c|c|}
\hline
\textbf{Video} & \textbf{\#} & \textbf{\#} & \textbf{\# Adj.} & \textbf{Inter.} & \multicolumn{1}{l|}{\textbf{Adj. inter.}} \\ 
& \textbf{Int.}    &  \textbf{Inter.} &   \textbf{inter.} & \textbf {(\%)} & \textbf{(\%)} \\ \hline
G2 T1          & 207                      & 26                        & 14                                 & 12                        & 6                                                       \\ \hline
G6 T1          & 157                      & 28                        & 20                                 & 17                        & 12                                                      \\ \hline
G110 T2        & 208                      & 52                        & 44                                 & 25                        & 21                                                      \\ \hline
\end{tabular}
\label{ICP}
\end{center}
\end{table}

\section{Experimental Results}

Experiments considered 16 teams. Two separate videos were processed for each team. Results were labeled as $Gxyz$ $Tk$, where label $xyz$ is the group identifier and label $k$ is $1$ or $2$ denoting the first or the second video for a group. Each video is more than 30 minutes long. 

\subsection{Speaker vs. time charts}


\begin{table}[t]
\caption{Training times of primary Neural Network models}
\begin{center}
	\begin{tabular}{|c|c|}
		\hline
		\textbf{Neural Network Model} & \textbf{Training Time (h:mm:ss)} \\ \hline
		Speaker Diarization           & 2:07:30                          \\ \hline
		SER                           & 2:29:59                          \\ \hline
	\end{tabular}
\label{NNT1}
\end{center}
\end{table}

As shown in Table~\ref{NNT1}, the training times of the primary NN models are about two hours long. These times do not include the duration required to form the training datasets. The SER model requires more data in fewer categories to successfully converge, which results in a longer training time. Speaker diarization model converges while requiring less data in more categories. Overall, the accuracy previously stated for each of these models justifies the training time required.

\begin{table*}[t]
\caption{Execution time of speaker diarization stages}
\begin{center}
	\begin{tabular}{|c|p{3cm}|p{3cm}|p{3cm}|p{3cm}|}
		\hline
		\textbf{Video} & \textbf{Data Segmentation (h:mm:ss)} & \textbf{CNN Predictions (h:mm:ss)} & \textbf{Spectral Clustering (h:mm:ss)} & \textbf{Temporal Smoothing (h:mm:ss)} \\ \hline
		G2 T1          & 0:00:01                              & 0:00:17                            & 0:28:21                                & 0:00:01                               \\ \hline
		G2 T2          & 0:00:01                              & 0:00:18                            & 0:36:52                                & 0:00:01                               \\ \hline
		G3 T1          & 0:00:02                              & 0:00:22                            & 1:10:02                                & 0:00:01                               \\ \hline
		G3 T2          & 0:00:02                              & 0:00:13                            & 0:08:01                                & 0:00:01                               \\ \hline
	\end{tabular}
\label{SD-ET}
\end{center}
\end{table*}

The execution time of each stage of speaker diarization is outlined in Table~\ref{SD-ET}. The data segmentation and temporal smoothing stages are the least computationally complex, and are executed almost instantaneously. The NN prediction execution time is linearly proportional to the amount of segmented data. The computational cost of this stage is not significant, as it takes tens of seconds to execute. The spectral clustering stage is the most computationally complex. Depending on the amount of segmented data, this stage can take minutes to hours to execute. This high execution time is due to the eigenvalue and eigenvector computation of the segmented data matrix. The only way to improve this execution time is to rewrite the numpy algorithm to be more efficient.

\begin{figure*}[t]
	\centering
	\begin{subfigure}[h]{0.4\textwidth}
		\centering
		\includegraphics[width=\textwidth]{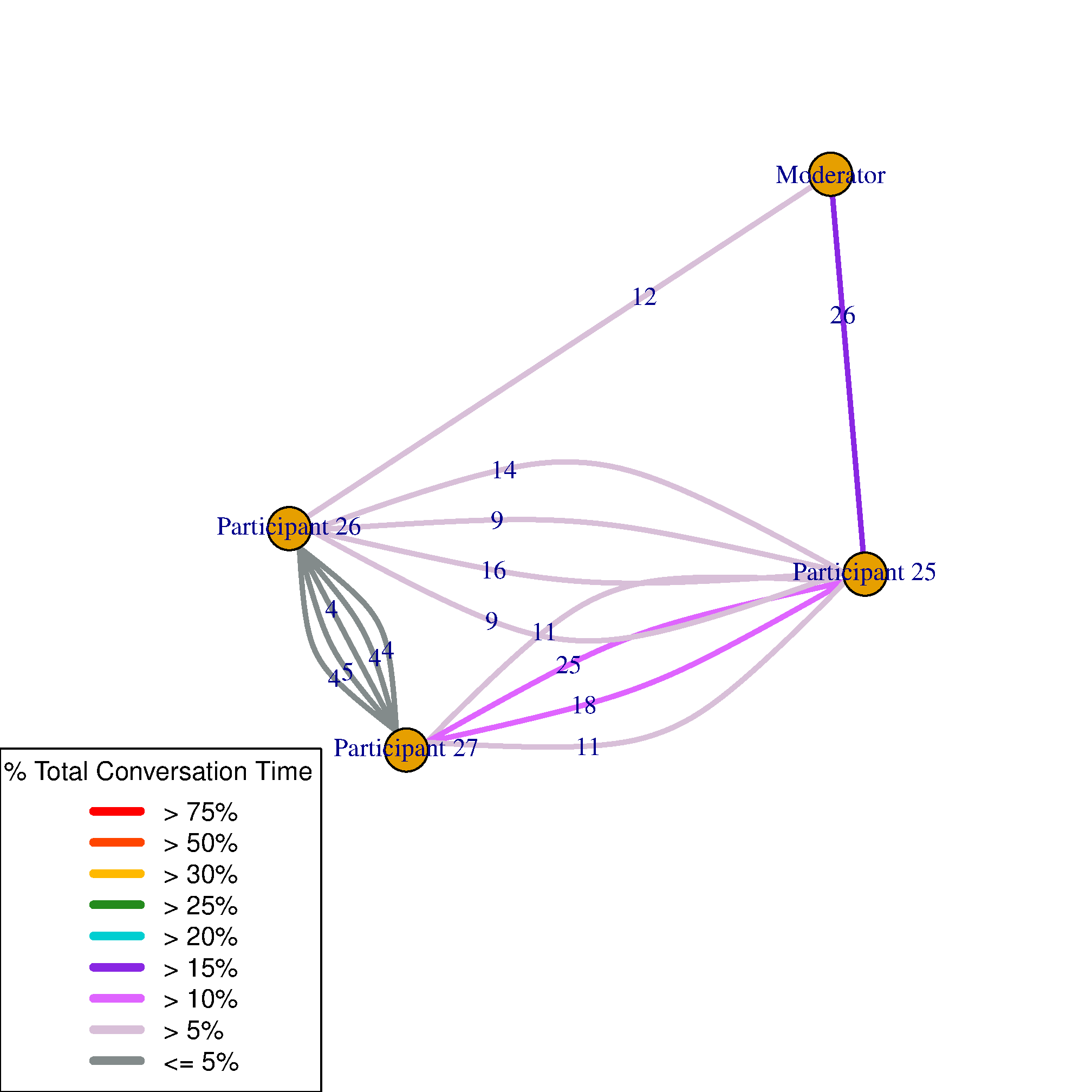}
		\caption{G7 T1}
		\label{SI1-1}
	\end{subfigure}
	\hfill
	\begin{subfigure}[h]{0.4\textwidth}
		\centering
		\includegraphics[width=\textwidth]{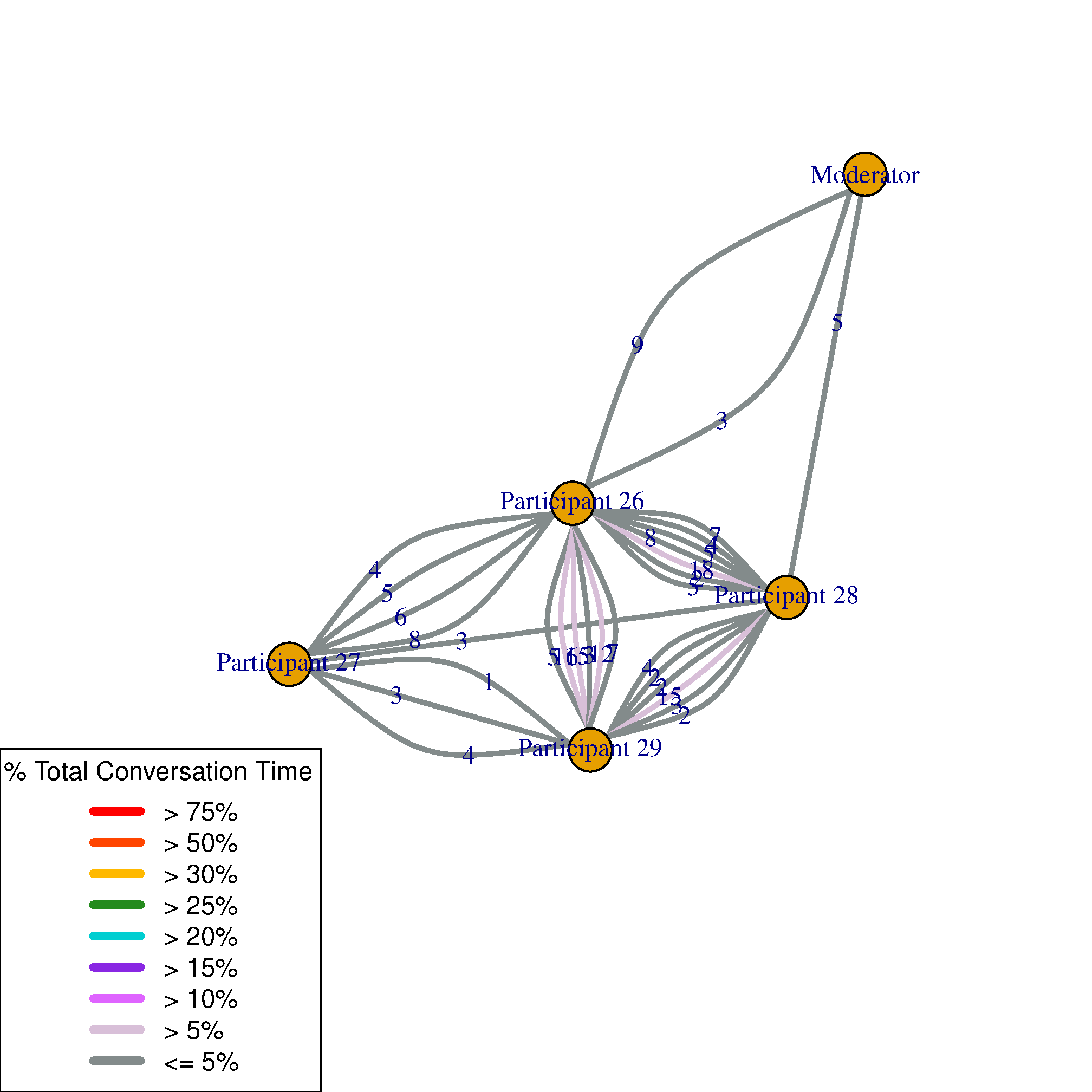}
		\caption{G7 T2}
		\label{SI1-2}
	\end{subfigure}
	\newline
	\begin{subfigure}[h]{0.4\textwidth}
		\centering
		\includegraphics[width=\textwidth]{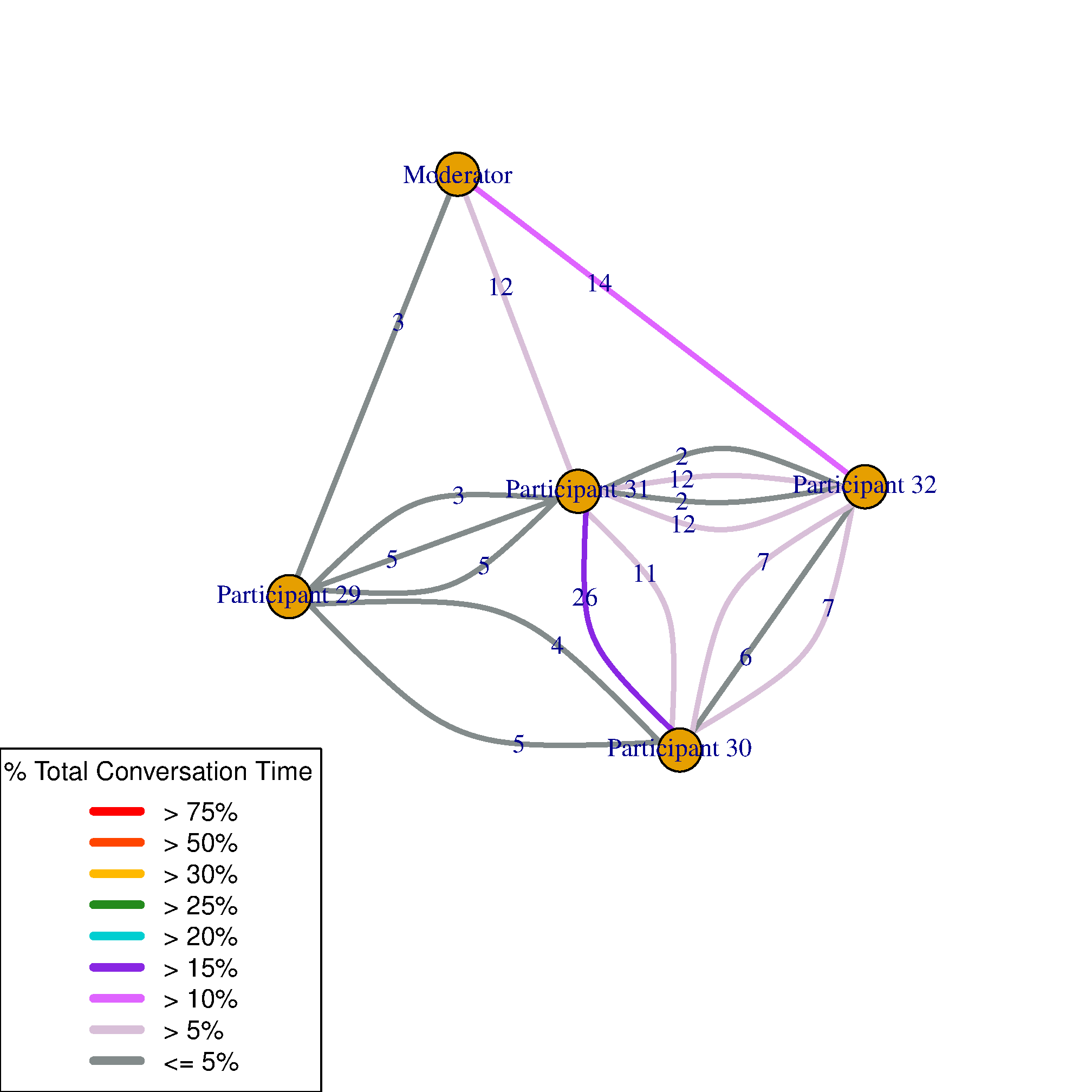}
		\caption{G8 T1}
		\label{SI1-3}
	\end{subfigure}
	\hfill
	\begin{subfigure}[h]{0.4\textwidth}
		\centering
		\includegraphics[width=\textwidth]{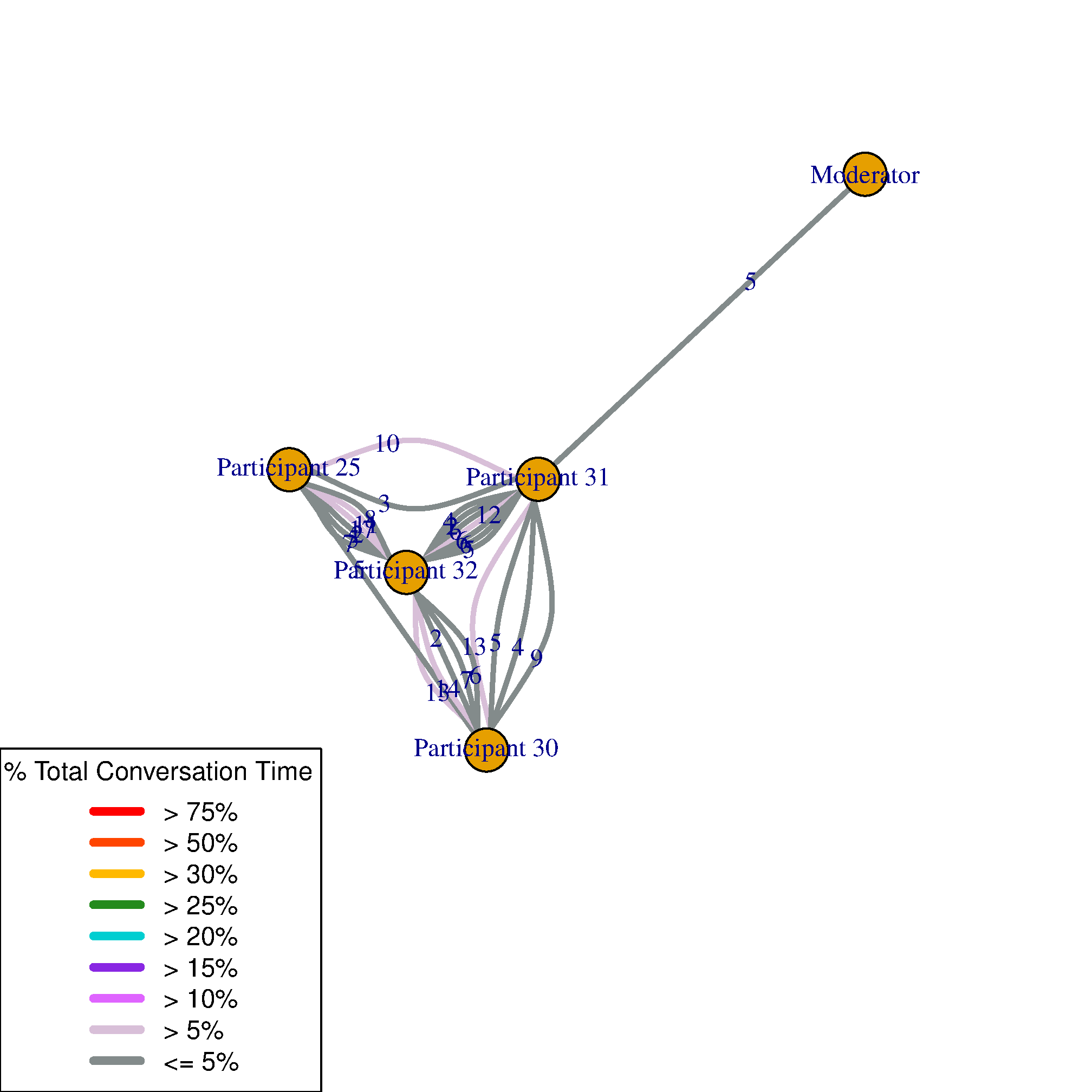}
		\caption{G8 T2}
		\label{SI1-4}
	\end{subfigure}
	\caption{First 2 minutes of interactions from videos G7 and G8}
	\label{SI1}
\end{figure*}

\subsubsection{Comparisons of speaker contributions within similar social situations}

Speaker diarization results are discussed in the paper for four videos from teams $G2$ and $G3$. The entire videos were processed, but only the first two minutes of speech were displayed in speaker vs. time charts. These charts are referenced in Figure~\ref{ST1}. 

As a general observation, the method has some drawbacks, as it is not always accurate. Within some videos, the assigned IDs of two speakers were swapped, but the rest were assigned correctly. The accuracy of this algorithm depends solely on the detection accuracy of the utterances derived through the speaker diarization process. 

The following conclusion was extracted about speaker trends using only data diarization results. 
Within videos G2 T1, G2 T2, and G3 T2, two speakers speak on a consistent basis. One additional speaker speaks on a less consistent basis, and the remaining speaker does not speak much at all. In the video of G3 T1, three speakers speak on a consistent basis, while the remaining speaker speaks less frequently. These properties indicate a similarity in speaker behavior. Participants 6, 7, and 8 are in both videos G2 T1 and G2 T2. However, Participant 6 speaks less in the G2 T1 video, but more in the G2 T2 video. Participants 7 and 8 feature similar dynamics in both videos G2 T1 and G2 T2. Similarly, participants 10, 11, and 12 are reused in videos G3 T1 and G3 T2. Participants 10 and 11 speak more in video G3 T1 than in video G3 T2. Participant~12 features similar dynamics for the two videos. While there is a change in behavior for Participant~6, attributing this change to the different participant in the two videos would false as in both videos the contribution of the fourth member is very small. Hence, while there were different participants in the two videos, there behavior was similar. Therefore, this example describes a situation in which hypotheses extraction concludes that the observed changes in team behavior cannot be related to the modified parameters, and a different analysis granularity must be considered. 


\begin{figure*}[t]
	\centering
	\begin{subfigure}[h]{0.4\textwidth}
		\centering
		\includegraphics[width=\textwidth]{G7_T1_Speaker_Interactions_1.png}
		\caption{\textit{T}: 0-2 minutes}
		\label{SI2-1}
	\end{subfigure}
	\hfill
	\begin{subfigure}[h]{0.4\textwidth}
		\centering
		\includegraphics[width=\textwidth]{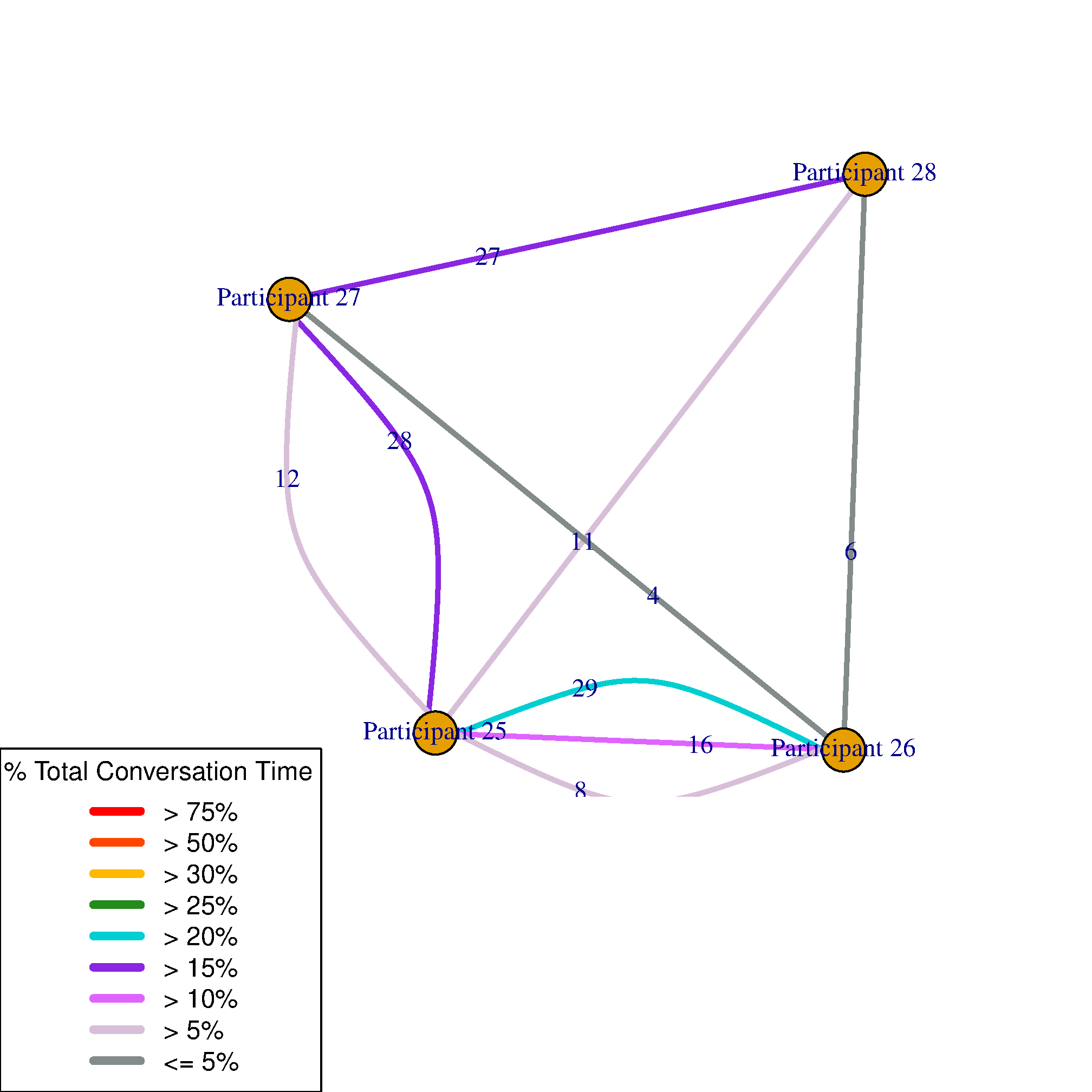}
		\caption{\textit{T}: 2-4 minutes}
		\label{SI2-2}
	\end{subfigure}
	\newline
	\begin{subfigure}[h]{0.4\textwidth}
		\centering
		\includegraphics[width=\textwidth]{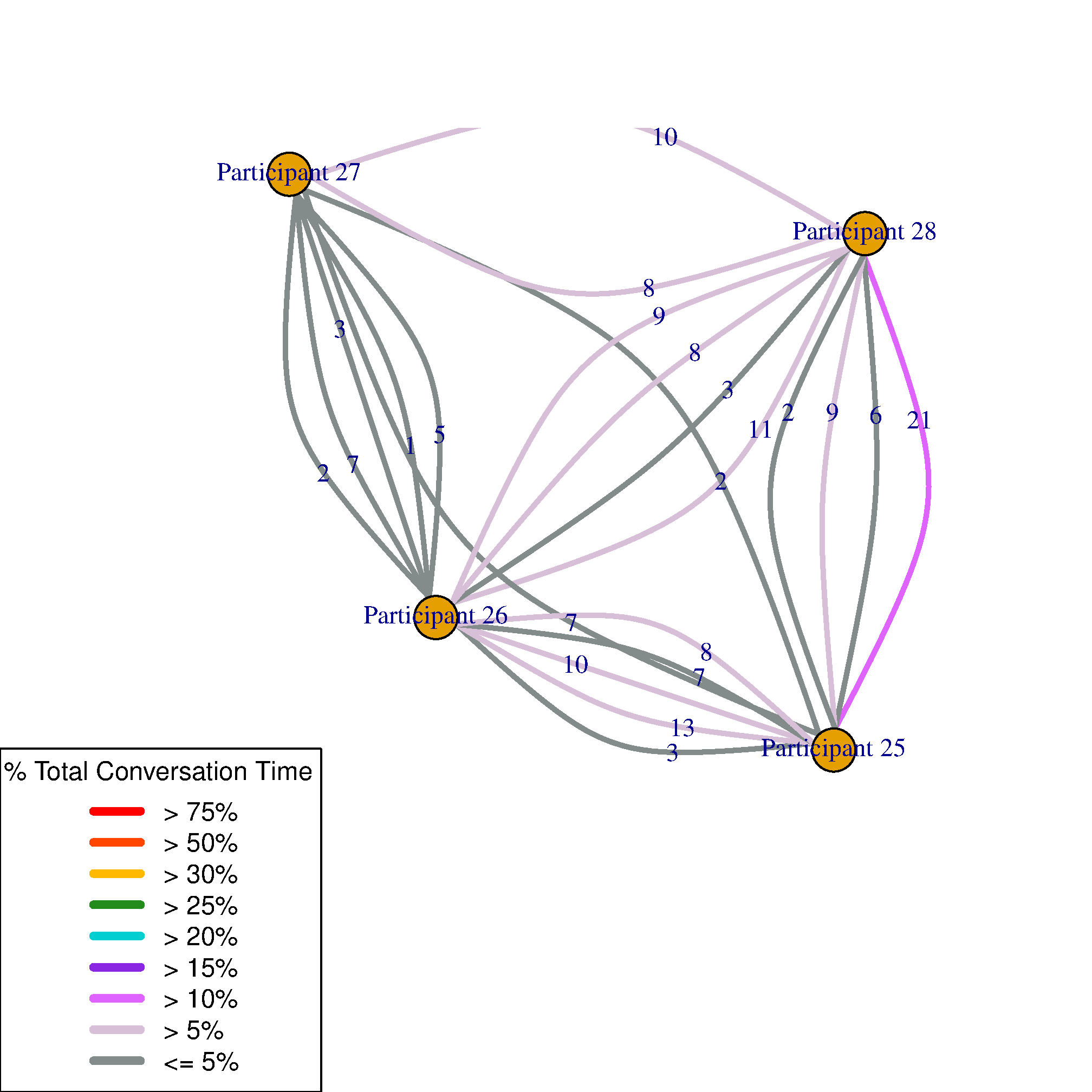}
		\caption{\textit{T}: 4-6 minutes}
		\label{SI2-3}
	\end{subfigure}
	\hfill
	\begin{subfigure}[h]{0.4\textwidth}
		\centering
		\includegraphics[width=\textwidth]{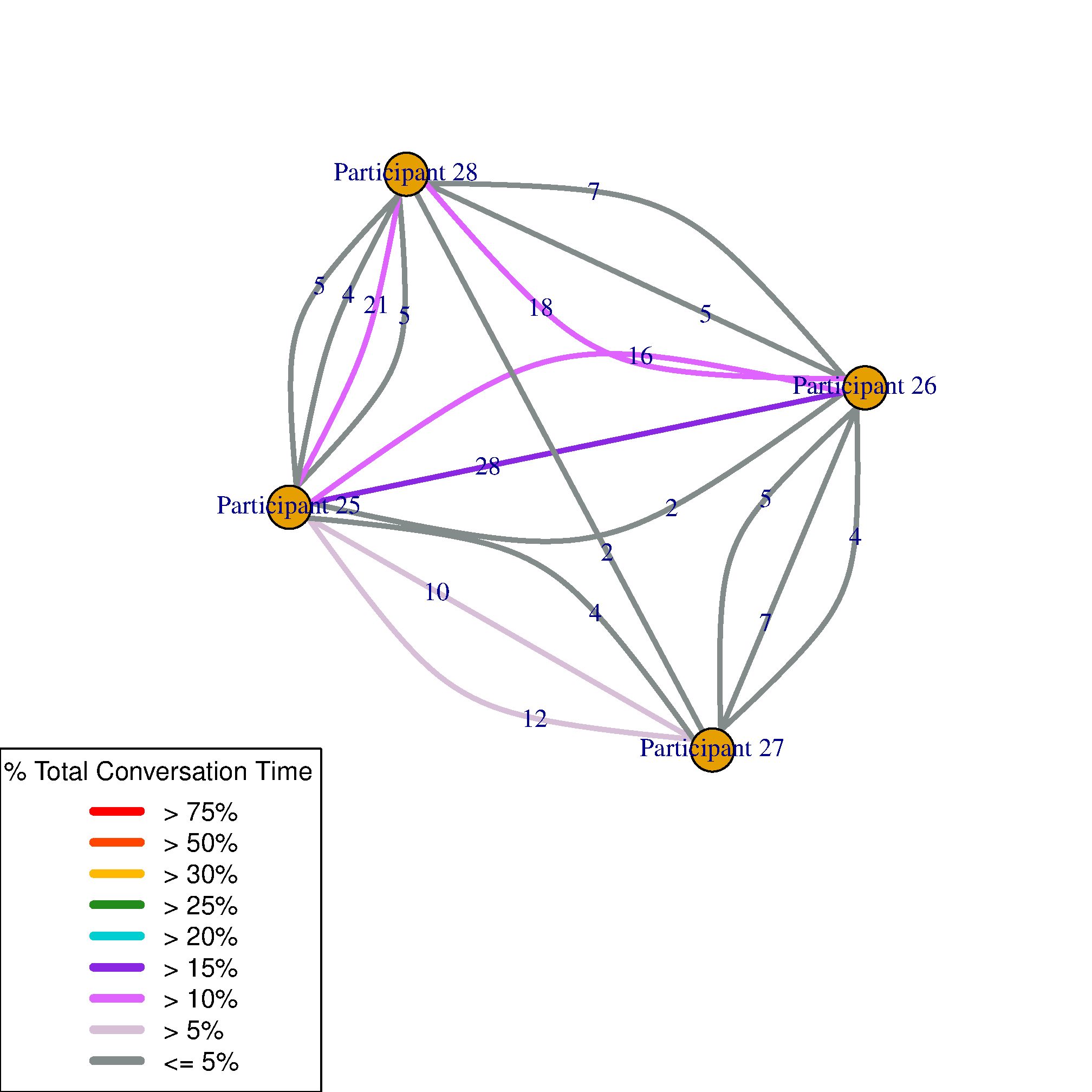}
		\caption{\textit{T}: 6-8 minutes}
		\label{SI2-4}
	\end{subfigure}
	\caption{Video G7 T1 Interaction Graphs (IGs)}
	\label{SI2}
\end{figure*}

\subsubsection{Speech trends over time}

Next, extracted hypotheses are discussed for speaker trends over time from various videos. These trends are evaluated within the first four two-minute segments of speech within each video, but similar observations exist for the remaining time of the videos.

For video G7 T1, varying speaker dynamics are present. These charts are displayed in Figure~\ref{ST2}. Participant 28 does not speak for the entire first two minutes of speech. This participant speaks less frequently in the second two minutes, but more in the third two minutes. In the fourth two minute segment, Participant 28 speaks less frequently. Participants 25, 26 and 27 feature similar speech patterns across the entire 8 minute interval. Participant 28's speech patterns indicate little contribution compared to the others.
For video G7 T2, a constant speaker dynamic is present. These charts are displayed in Figure~\ref{ST3}. Participants 26, 27, 28, and 29 present similar speech patterns across the entire 8 minute interval. Therefore, all participants contributed equally to the conversation. 

Similarity and dissimilarity analysis between the trends in the two videos suggests that the change in the behavior of Participant~28 could relate to replacing an active participant, e.g., Participant~25, with a less active member, like Participant~25. It allowed Participant~28 to become engaged in the discussion. The extracted hypothesis describes this observation.   




\begin{figure*}[t]
	\centering
	\begin{subfigure}[h]{0.4\textwidth}
		\centering
		\includegraphics[width=\textwidth]{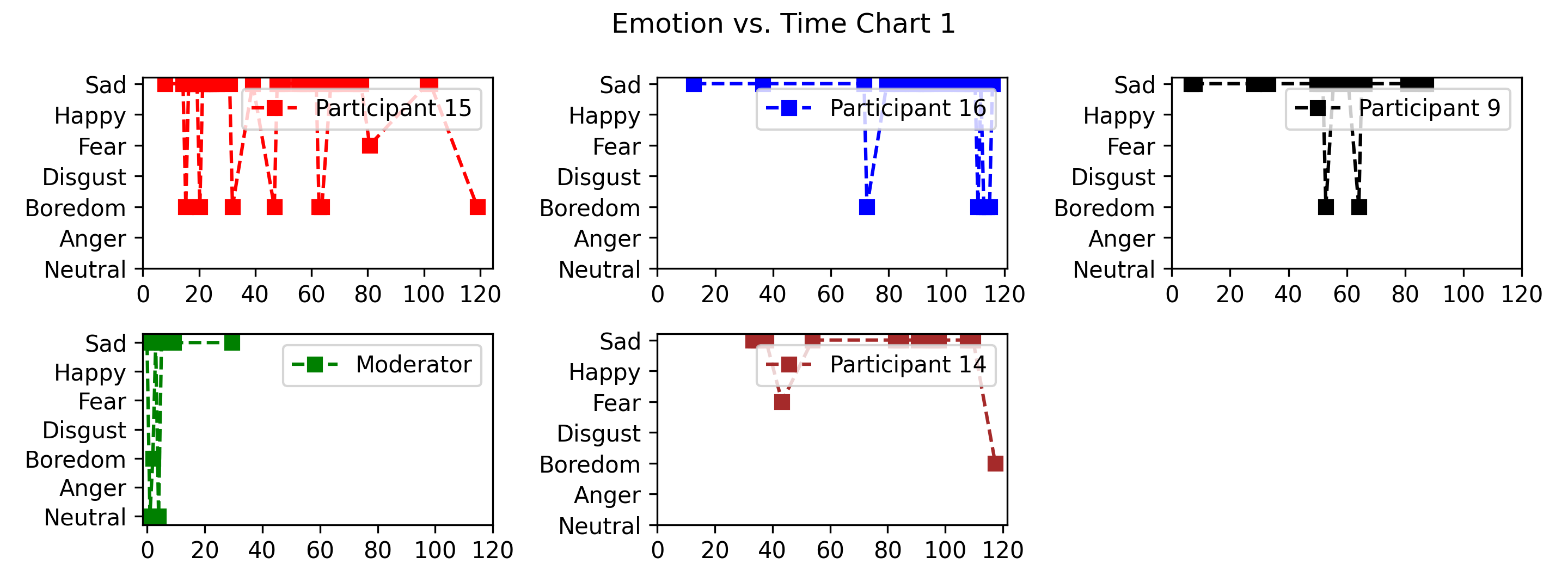}
		\caption{\textit{T}: 0-2 minutes}
		\label{SE2-1}
	\end{subfigure}
	\hfill
	\begin{subfigure}[h]{0.4\textwidth}
		\centering
		\includegraphics[width=\textwidth]{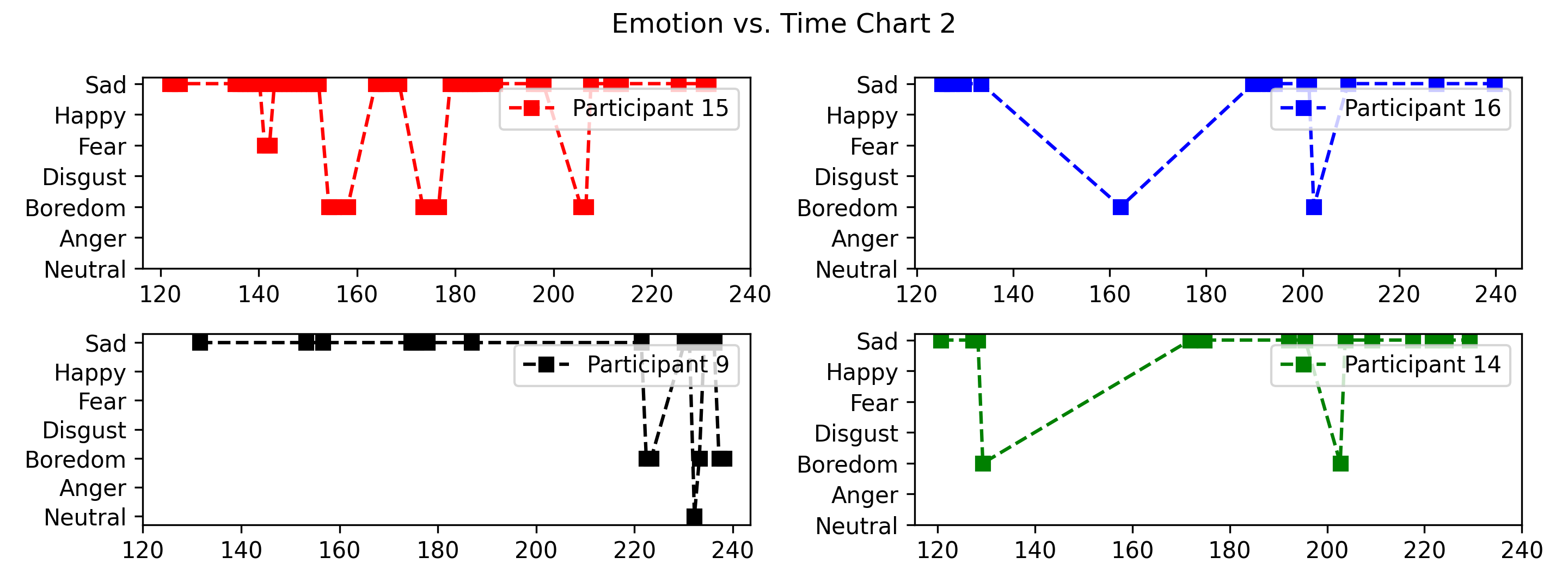}
		\caption{\textit{T}: 2-4 minutes}
		\label{SE2-2}
	\end{subfigure}
	\newline
	\begin{subfigure}[h]{0.4\textwidth}
		\centering
		\includegraphics[width=\textwidth]{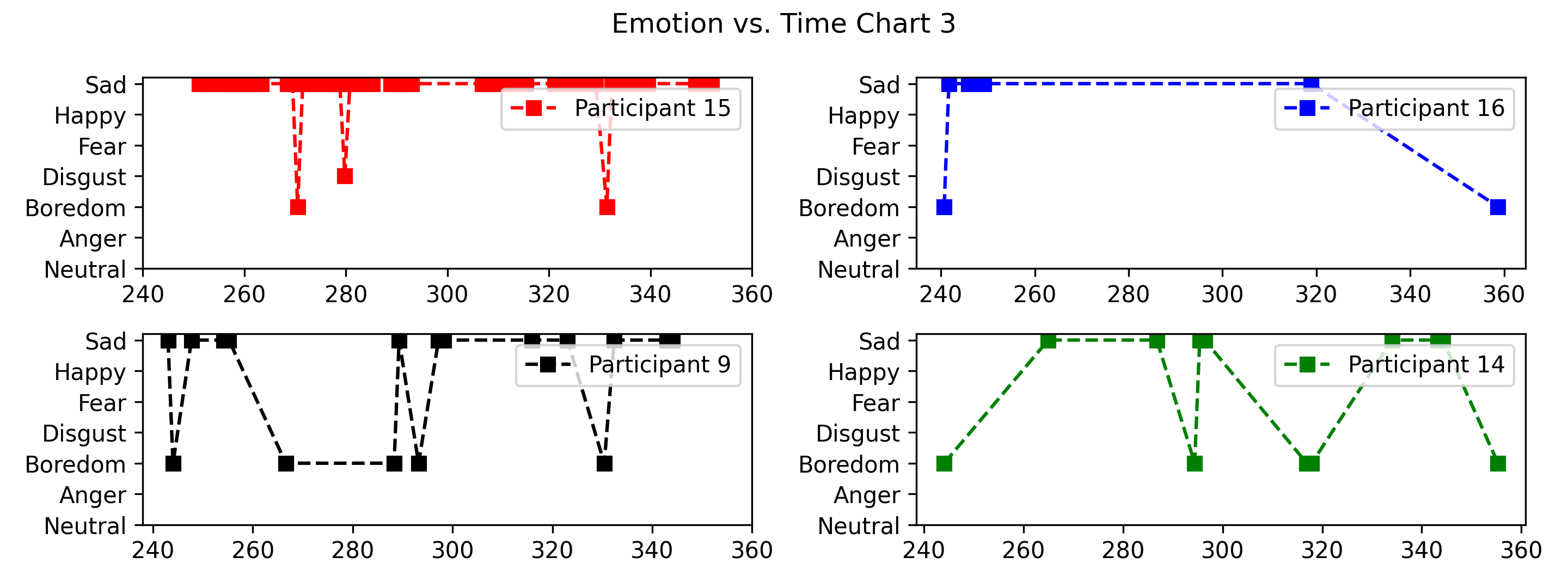}
		\caption{\textit{T}: 4-6 minutes}
		\label{SE2-3}
	\end{subfigure}
	\hfill
	\begin{subfigure}[h]{0.4\textwidth}
		\centering
		\includegraphics[width=\textwidth]{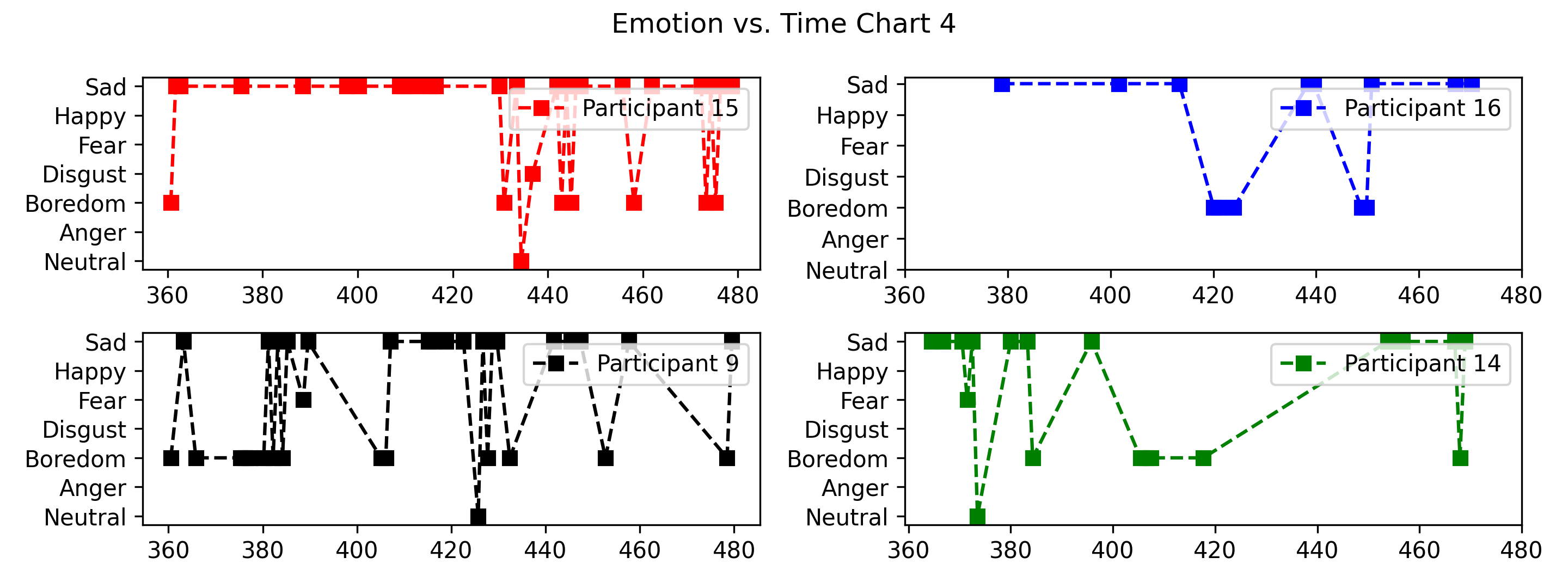}
		\caption{\textit{T}: 6-8 minutes}
		\label{SE2-4}
	\end{subfigure}
	\caption{Video G4 T2 emotion trends}
	\label{SE2}
\end{figure*}

\subsection{Speaker Interaction Graphs (IGs)}

The computational complexity of the interaction characterization algorithm is linearly proportional to the number of utterances detected by Speaker diarization. This algorithm executes within tens of seconds. Therefore, it has a negligible effect on the execution time of the system.

Table~\ref{ICP} summarizes the performance of speaker interaction detection, specifically, the frequency of interruptions between speakers. Results are shown for three videos. As expected, interruptions can lead to interaction characterization errors. The percentage of interruptions in each video were 12\%, 17\%, and 25\% respectively. 
However, interruptions only affect the algorithm performance if they disrupt the normal conversation flow of the video. By adjusting our metric to include only those interruptions, the percentage interruptions for each video are reduced to 6\%, 12\%, and 21\%. It is also important to note that the majority of the adjusted interruptions occur within short bursts of interactions, in which speakers interject and speak very frequently within a short amount of time. Therefore, they can be considered to not contribute significantly to the topic of the conversation. Therefore, the accuracy of the interaction algorithm is adequate despite these anomalies.

IGs offer a different facet of speaker participation and interactions than speech diarization charts. They highlight emerging self-organization (self-structuring) of team interactions, like participant clustering, dominant members, and outliers. Hence, the extracted hypotheses describe any invariant aspects of team interaction self-organization. Speaker IGs are discussed for four videos for teams $G7$ and $G8$.

\subsubsection{Comparisons of speaker interactions within similar social situations}

The first two minutes of team interaction are displayed in the IGs in Figure~\ref{SI1}. 
Within videos G7 T2 and G8 T2, the interactions form in a central web between the four participants, while the moderator exists in an interaction branch outside of the main web. For videos G7 T1 and G8 T1, the interactions exist within a web of all five speakers. An extracted hypothesis suggests that it is unlikely that within the first two minutes all participants equally contribute to the discussions. Instead, it is likely that one participant is more active or discussions are in a subgroup. 


\subsubsection{Interaction trends over time}

Hypotheses extraction considered interaction trends over time from various videos. Next, trends within the first four two-minute segments of speech within each video are discussed.

\paragraph{Videos G7 T1-G7 T2 and G8 T1-G8 T2}

Video G7 T1 displays a constant interaction dynamic between the four participants within the video. These charts are shown in Figure \ref{SI2}. Participants 25, 26, 27, and 28 are represented as an interaction web with no outliers. 
The same interaction dynamic exists for video G7 T2 too, 
which includes team members 26, 27, 28, and 29 (team members 25 and 29 were swapped).
In contrast, video G8 T1 displays a variable interaction dynamic between the four participants within the video. 
Participants 29, 30, 31, and 32 are represented as an interaction web with no outliers in the first, third, and fourth time intervals. Participants 30, 31, and 32 had singular interactions between each other in the second time interval, with Participant 29 missing from the conversation. 
Then, video G8 T2 shows a constant interaction dynamic between the four participants within the video. 
Participants 25, 30, 31, and 32 are represented as an interaction web with no outliers. 
The extracted hypotheses suggests that the interaction behavior of the three members that are in both videos T1 and T2 does not change because of swapping one team member. However, the behavior of the swapped member can be different in the two teams. However, it is unclear what parameters of the interaction of the three members influence the interaction of the swapped participant. 


\subsection{Emotion vs. time charts}

Tracking emotion trends over time from various videos are discussed next. These trends correspond to the first four two-minute segments of speech within each video. Each evaluation involves measuring the amount of deviations from the algorithmic fallback emotion.

\begin{table*}[t]
\caption{Execution time of SER stages}
\begin{center}
	\begin{tabular}{|c|p{3cm}|p{3cm}|p{3cm}|}
		\hline
		\textbf{Video} & \textbf{Data Segmentation (h:mm:ss)} & \textbf{CNN Predictions (h:mm:ss)} & \textbf{Data Concatenation (h:mm:ss)} \\ \hline
		G2 T1          & 0:00:08                              & 0:03:25                            & 0:00:12                               \\ \hline
		G2 T2          & 0:00:10                              & 0:04:21                            & 0:00:13                               \\ \hline
		G3 T1          & 0:00:14                              & 0:05:19                            & 0:00:17                               \\ \hline
		G3 T2          & 0:00:10                              & 0:02:15                            & 0:00:09                               \\ \hline
	\end{tabular}
\label{SER-ET}
\end{center}
\end{table*}

The execution time of each stage of the SER process is outlined in Table \ref{SER-ET}. The data segmentation and data concatenation stages are the least computationally complex, and are executed within tens of seconds. The NN prediction stage is linearly proportional to the amount of segmented data. However, the computational cost is less than the execution time of the spectral clustering stage within the previous module. Therefore, CNN predictions of SER has an insignificant contribution to the overall execution time of the diaLogic system.

\paragraph{Videos G4 T1 and G4 T2}

Video G4 T1 presents a changing emotional dynamic. These charts are depicted in Figure~\ref{SE1}. Participants 13, 15, and 16 feature a similar number of emotional deviations from \textit{Sad} across the entire 8~minute interval. Participant~14 has a similar number of emotional deviations from \textit{Sad} in the first, second, and fourth time intervals, but had more deviations in the third time interval. 
Video G4 T2 also has a changing but different emotional dynamic a displayed in Figure~\ref{SE2}. Participant 15 experiences a higher number of emotional deviations from \textit{Sad} in the first and fourth time intervals, but less in the second and third time intervals. Participant 16 features a consistent number of deviations across the entire 8 minute interval. Participant 9 has less deviations within the first and second time intervals, but more in the third and fourth intervals. Participant 14 features incrementally more deviations as time increases. 

\paragraph{Videos G6 T1 and G6 T2}


Video G6 T1 exhibits a changing emotional dynamic. 
Participant~21 features a lower number of emotional deviations from \textit{Sad} in the first interval, but more in the fourth. Additionally, this participant yielded an unnatural number of deviations within the second and third time intervals. This unnatural level of deviation usually only occurs in the presence of significant background noise. Participant~22 has a consistent number of deviations across the entire 8 minute interval. Participant 23 experiences less deviations within the first and second time intervals, zero deviations within the third time interval, and more in the fourth interval. Participant 24 presents more deviations in the first and second time intervals, less in the fourth time interval, and zero deviations in the third interval. 
Video G6 T2 also presents a changing emotional dynamic. 
Participant 17 features zero deviations from \textit{Sad} in the first time interval, followed by an incrementally increasing number of deviations across the remaining three intervals. Participants~22 and 24 have an incrementally increasing number of emotional deviations from \textit{Sad} across the entire 8 minute interval. Participant 23 shows more deviations within the first time interval, less deviations within the second and fourth time intervals, and fewer still within the third time interval. 

The hypothesis extracted for the for videos suggests that the levels of emotional interactions in the team are correlated, as a member's changing increased emotional dynamic is preceded by another member's increase. 
This case suggests the existence of emotional coupling between participants.

\begin{table}[t]
\caption{Performance evaluation of Azure Speech-to-text}
\begin{center}
\begin{tabular}{|c|c|c|c|c|c|}
\hline
\textbf{Video} & \textbf{\#} & \textbf{\# Senten.} & \textbf{Senten.w/} & \textbf{\# Blank} & \textbf{Blank} \\ 
& \textbf{senten.} & \textbf{w/ error} & \textbf{error (\%)} & \textbf{senten.} & \textbf{senten. (\%)}\\ \hline
G2 T1          & 379                   & 17                             & 4                              & 18                          & 4                           \\ \hline
G3 T1          & 571                   & 44                             & 7                              & 17                          & 2                           \\ \hline
G4 T1          & 314                   & 19                             & 6                              & 20                          & 6                           \\ \hline
G6 T1          & 249                   & 11                             & 4                              & 12                          & 4                           \\ \hline
\end{tabular}
\label{STP}
\end{center}
\end{table}

\subsection{Speech-to-text conversion and words per minute estimation}

\begin{table}[t]
\caption{Average number of words per minute for video G2}
\centering
	\begin{tabular}{|c|c|c|}
		\hline
		\multicolumn{1}{|l|}{\textbf{Speaker}} & \multicolumn{1}{l|}{\textbf{Video}} & \multicolumn{1}{l|}{\textbf{Average \# words per min.}} \\ \hline
		Participant 1                               & G2 T2                               & 201                                                    \\ \hline
		Participant 5                               & G2 T1                               & 200                                                    \\ \hline
		Participant 6                               & G2 T1                               & 159                                                    \\ \hline
		Participant 6                               & G2 T2                               & 146                                                    \\ \hline
		Participant 7                               & G2 T1                               & 114                                                    \\ \hline
		Participant 7                               & G2 T2                               & 181                                                    \\ \hline
		Participant 8                               & G2 T1                               & 174                                                    \\ \hline
		Participant 8                               & G2 T2                               & 167                                                    \\ \hline
	\end{tabular}
\label{WPM}
\end{table}

\begin{table*}[t]
\caption{Excerpt of detected speech clauses from video G2 T1}
\centering
	\begin{tabular}{|p{7cm}|p{7cm}|}
		\hline
		\textbf{Sentence} & \textbf{Speech Clauses}                                                           \\ \hline
		I got one & Who: I, Consequences: got I                                                       \\ \hline
		Most effective bureaucracy ever. You can use an email and they respond within 14 days.                                                 & What: bureaucracy, Why: Because they use You, Consequences: use You, respond they \\ \hline
		Comma no other branch of the government works that effectively.                                                                        & What: government, How: no, Consequences: works government                         \\ \hline
		But even though they legalized it, they make you register as a narco                                                                   & Who: they, What, narco, For Who, you, Consequences: legalized they, register you  \\ \hline
		In reality, this is our solution.                                                                                                      & What: reality, Consequences: is reality                                           \\ \hline
		You don't have to register with the DEA                                                                                                & Who: You, What: DEA, How: You, Consequences: do You                               \\ \hline
		Register with the DEA. But you do have to register as a narcotic or dangerous drug manufacturer or distributor, so I wonder how. Legal & Who: you, What: DEA, For Who: I, How: the, But, register, Register with           \\ \hline
		I don't know some of the rules that they have                                                                                          & Who: I, What: rules, For Who: they, How: I, Consequences: do I, have rules        \\ \hline
		It's like they threw this together so quick.                                                                                           & For Who: they                                                                     \\ \hline
	\end{tabular}
\label{SCT}
\end{table*}

\begin{figure*}[t]
	\centering
	\begin{subfigure}[t]{0.3\textwidth}
		\centering
		\includegraphics[width=\textwidth]{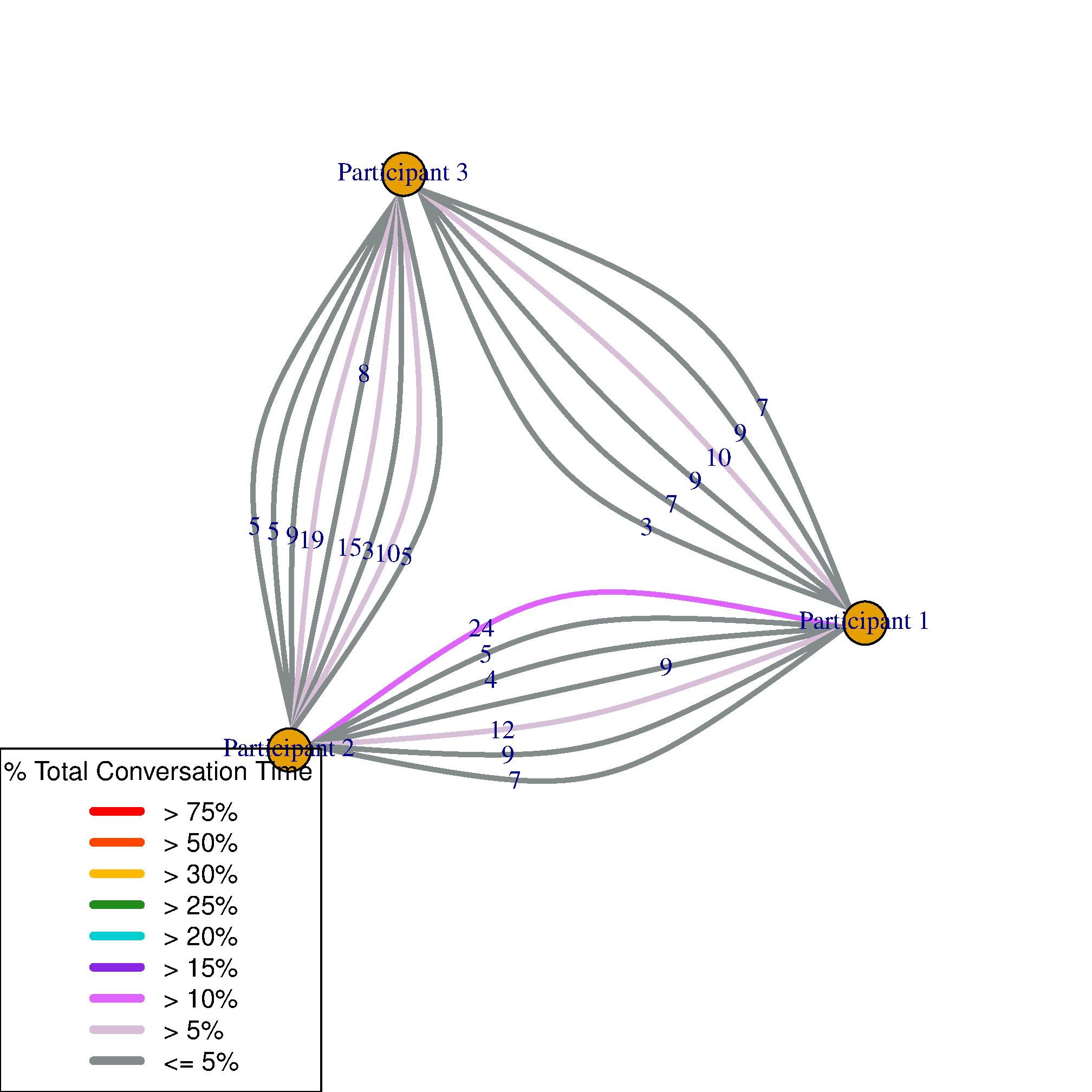}
		\caption{$T_1$ participant interactions 0-2 minutes}
		\label{C-1b-1}
	\end{subfigure}
	\hfill
	\begin{subfigure}[t]{0.3\textwidth}
		\centering
		\includegraphics[width=\textwidth]{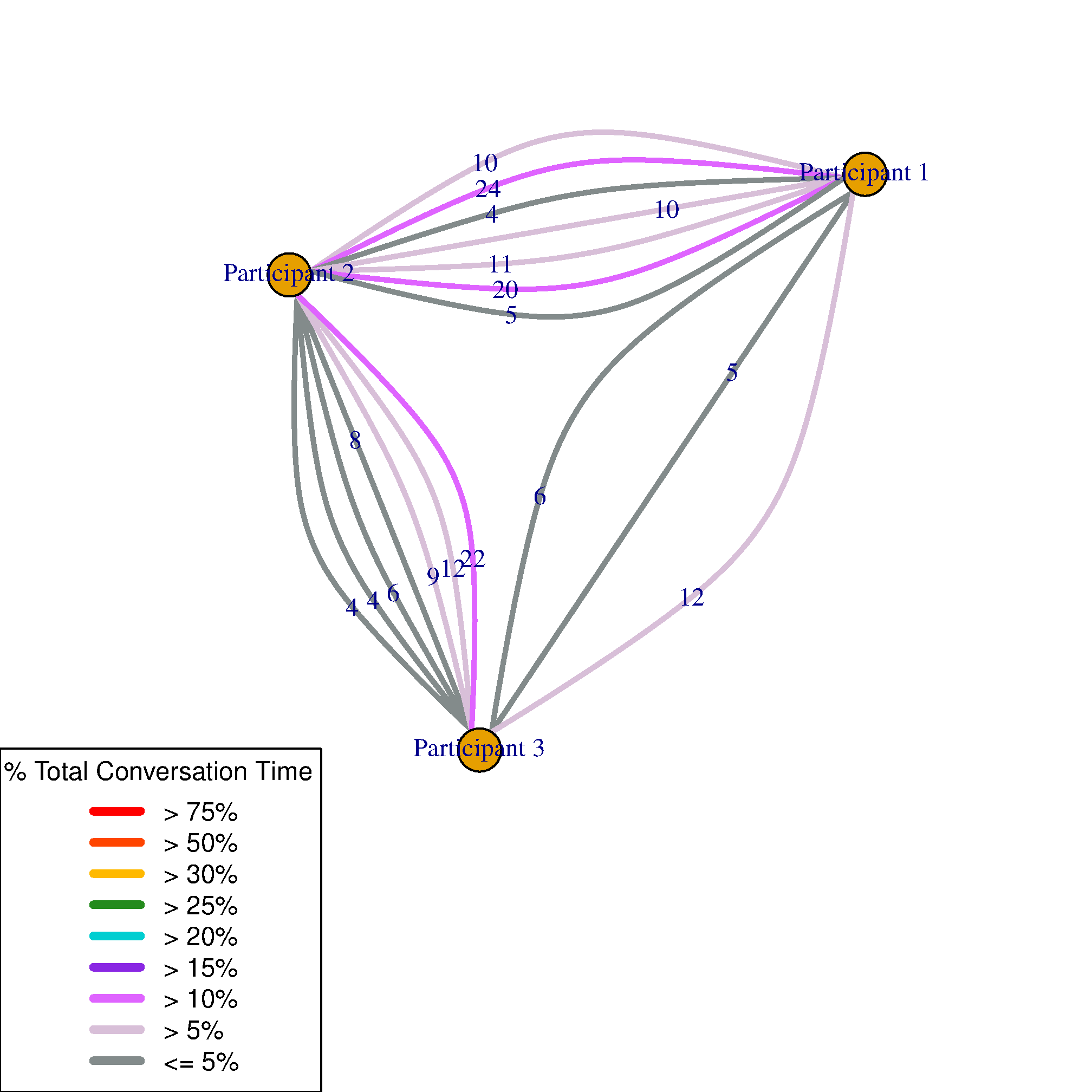}
		\caption{$T_2$ participant interactions 2-4 minutes}
		\label{C-1b-2}
	\end{subfigure}
	\hfill
	\begin{subfigure}[t]{0.3\textwidth}
		\centering
		\includegraphics[width=\textwidth]{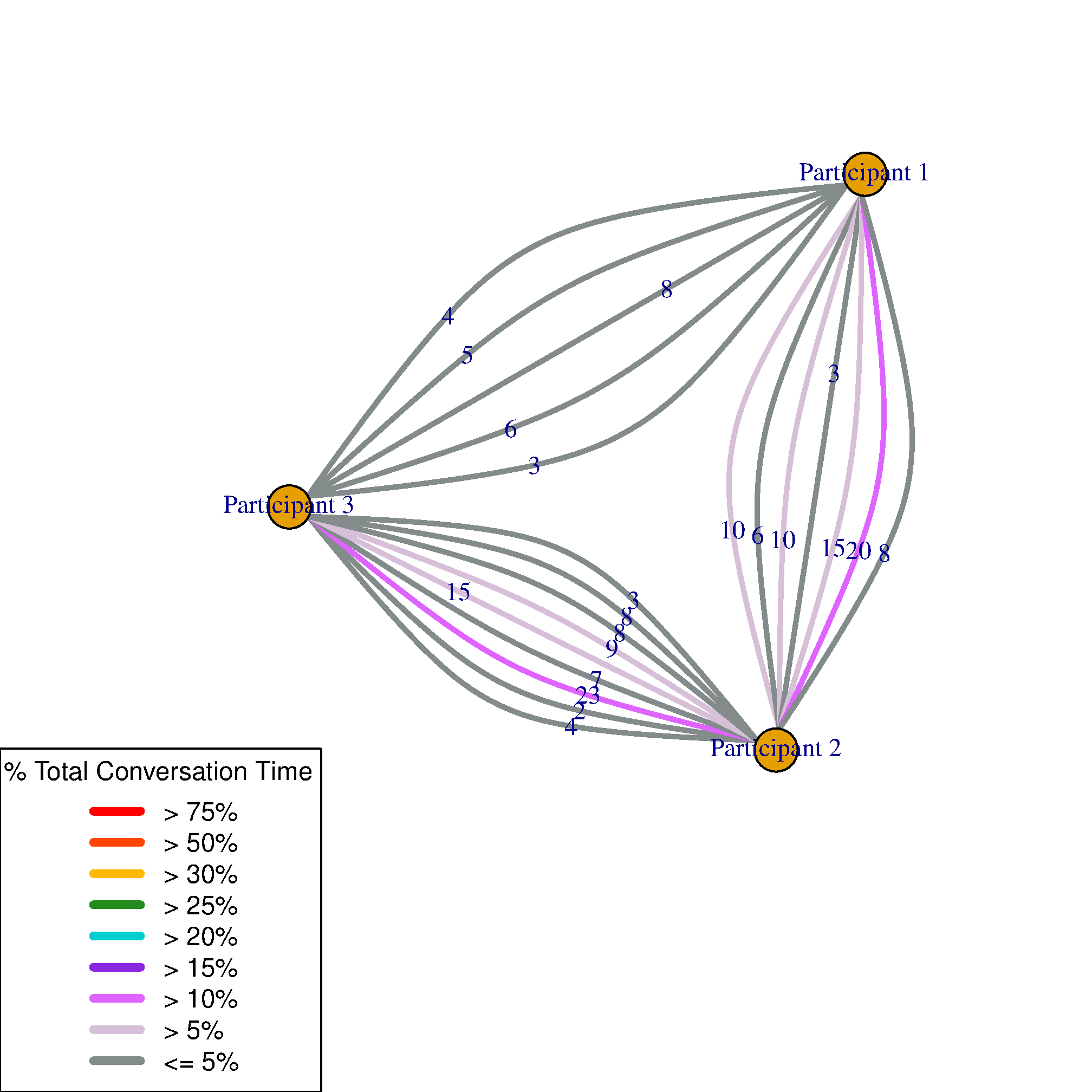}
		\caption{$T_3$ participant interactions 4-6 minutes}
		\label{C-1b-3}
	\end{subfigure}
	\newline
	\begin{subfigure}[t]{0.3\textwidth}
		\centering
		\includegraphics[width=\textwidth]{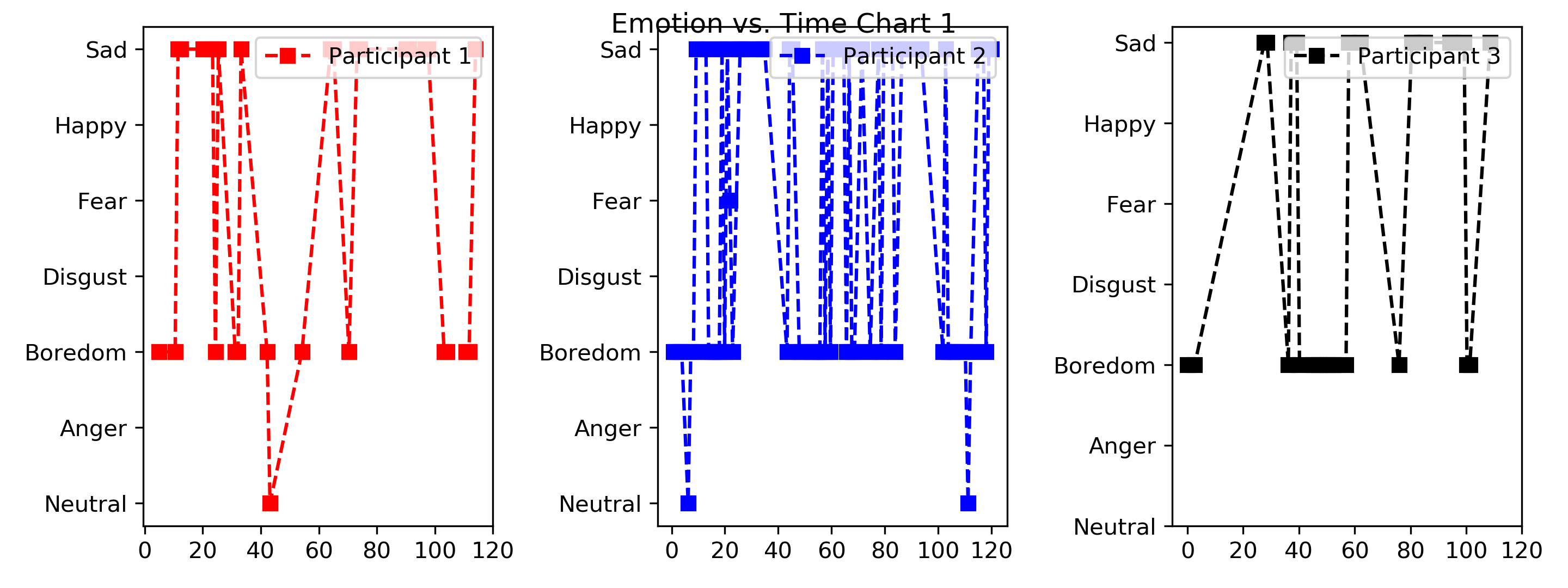}
		\caption{$T_1$ participant emotions 0-2 minutes}
		\label{C-1a-1}
	\end{subfigure}
	\hfill
	\begin{subfigure}[t]{0.3\textwidth}
		\centering
		\includegraphics[width=\textwidth]{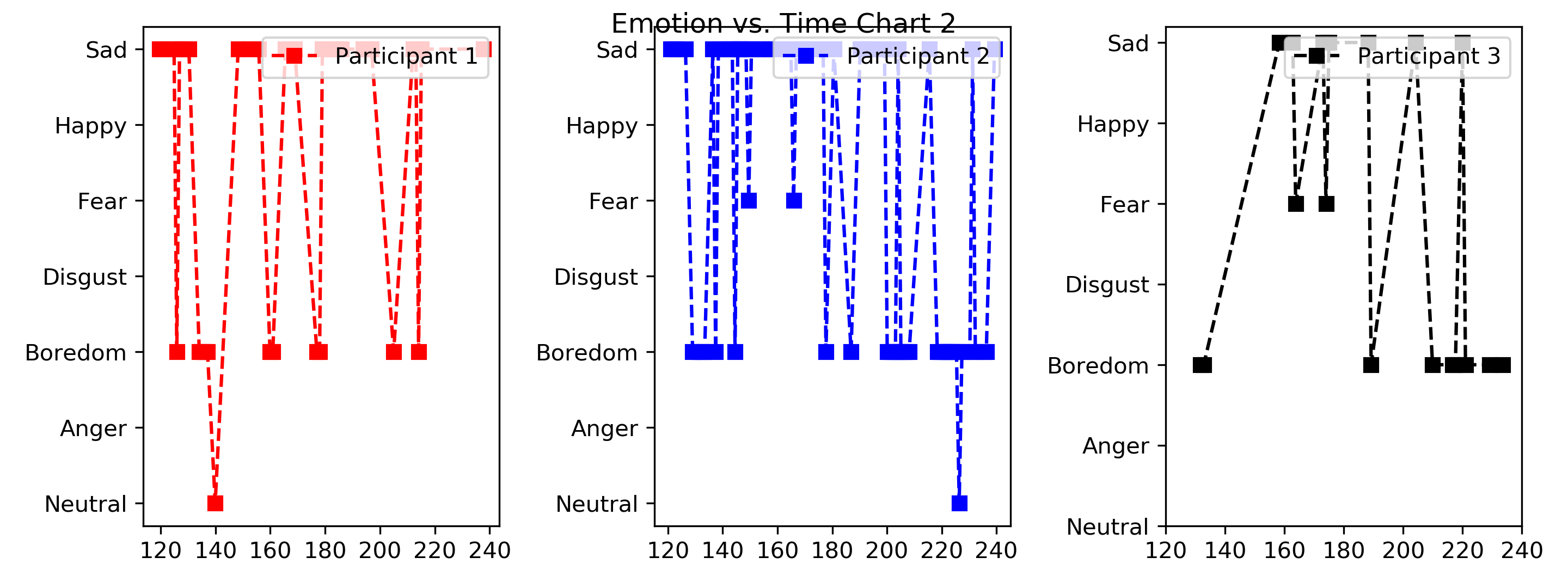}
		\caption{$T_2$ participant emotions 2-4 minutes}
		\label{C-1a-2}
	\end{subfigure}
	\hfill
	\begin{subfigure}[t]{0.3\textwidth}
		\centering
		\includegraphics[width=\textwidth]{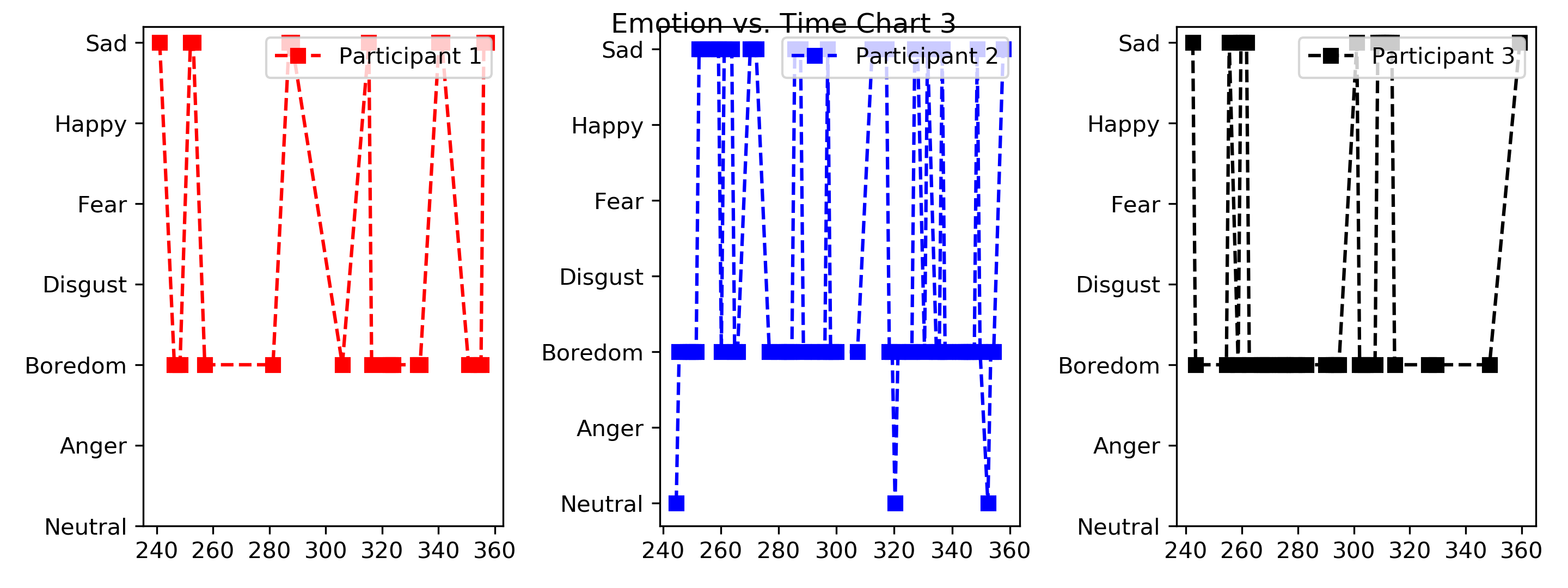}
		\caption{$T_3$ participant emotions 4-6 minutes}
		\label{C-1a-3}
	\end{subfigure}
	\caption{Constant interaction dynamic}
	\label{C-1}
\end{figure*}


The computational complexity of Azure speech-to-text algorithm is linearly proportional to the number of utterances. The execution time takes tens of minutes due to the repeated calls to Azure cloud platform. Overall, the speech-to-text algorithm features the highest cost on the total execution time of the diaLogic system except the NN training time.

Table~\ref{STP} presents the performance of the speech-to-text conversion for four videos. The most common errors were incorrect words and blank sentences. Incorrect words are immediately identifiable as being out of context for a specific sentence. Blank sentences can occur within speech segments of any length. 
The percentage of sentences with incorrect words within the four videos was 4\%, 7\%, 6\%, and 4\% respectively. The percentage of blank sentences was 4\%, 2\%, 6\%, and 4\% respectively. These results indicate a less than 10\% error for all cases, which underlines the robustness of Azure algorithm.

While the Azure speech library is superior to other speech-to-text algorithms, it featured some limitations. For some utterances, no text was detected. Furthermore, a period can be randomly inserted into a sentence. It censors obscenities, yet in some cases obscenities were detected where there weren't any. Some additional misspellings or incorrect words were also inserted into sentences. These inaccuracies can sometimes have an impact on the next NLP module of the system. Therefore, speaker diarization and interaction characterization are a more accurate representation of group performance. 

Speech-to-text results are shown for videos from team~G2. The average words-per-minute rate for each participant are listed in Table \ref{WPM}. Members 6, 7, and 8 participated in both videos T1 and T2. Participant 6 and Participant 8 yielded a similar speech rate across the two videos. Participant 7 featured a speech rate reduction from video T1 to video T2. Hence, it can be hypothesized that Participant~7 contributed less within video T2 as opposed to video T1. However, these results do not represent interactions as a whole. 
These rates are only a facet of a specific participant's performance. 

\begin{table}[t]
\caption{Performance evaluation of speech clause detection algorithm}
\begin{center}
\begin{tabular}{|c|c|c|c|c|c|}
\hline
\textbf{Video} & \textbf{\#} & \textbf{\# Senten.} & \textbf{Senten.} & \textbf{\# Senten.} & \textbf{Senten.} \\ 
& \textbf{Senten.} & \textbf{w/ambig.} & \textbf{w/ambig.} & \textbf{w/alg.} & \textbf{w/alg.} \\ 
&  & \textbf{error} & \textbf{error(\%)} & \textbf{error} & \textbf{error(\%)} \\ \hline
G7 T1          & 50                    & 33                                       & 66                                       & 12                                       & 24                                       \\ \hline
G8 T1          & 50                    & 37                                       & 74                                       & 9                                        & 18                                       \\ \hline
G9 T1          & 50                    & 34                                       & 68                                       & 6                                        & 12                                       \\ \hline
G10 T1         & 50                    & 27                                       & 54                                       & 8                                        & 16                                       \\ \hline
\end{tabular}
\label{SCP}
\end{center}
\end{table}

\subsection{Speech clauses detection}

The computational complexity of the speech clauses algorithm is directly proportional to the number of sentences within the entire set of detected speech from the speech-to-text algorithm. The most costly components of this algorithm are the repeated calls to CoreNLP and pyWSD, both which have cloud components. The execution time of this algorithm is in the singular minutes range.

Table~\ref{SCP} summarizes the performance of the clause detection algorithm for the first 50 speech clauses results from four videos. The most common anomalies were ambiguity determination errors within CoreNLP and PyWSD, and errors within the main algorithm. The most common ambiguity errors occur within parts of speech detection, where a specific word is not correctly determined to be a noun, verb, adjective, or adverb. Furthermore, nouns which correspond to categories of \textit{PERSON}, \textit{ORGANIZATION}, \textit{MISC}, {DATE, TIME, DURATION}, \textit{SET}, or \textit{LOCATION} are not detected as such or are mislabeled. The most common algorithm errors occur when a noun in the context of \textit{For Who} is stated at the start of a sentence, incorrectly marking it as \textit{Who}, or an entity which is represented as \textit{What}, \textit{When}, or \textit{Where} occurs within multiple words. The algorithm assumes that each clause is a single word, which affects the detection accuracy. 
The percentage of ambiguity errors for the first 50 speech clauses within the four videos was 66\%, 74\%, 68\%, and 54\% respectively. The percentage of algorithm errors was 24\%, 18\%, 12\% and 16\%. These results show that the detection accuracy of the algorithm as a whole is being held back significantly by the ambiguity handling of CoreNLP and PyWSD. The algorithm itself features an adequate accuracy. Further improvements to ambiguity handling may even improve the algorithm accuracy. 

Speech clause detection was shown on the video G2 T1. Some of the results are displayed in Table~\ref{SCT}. As shown, sentences which feature general language formats yield the most accurate detection. Furthermore, the rule-based approach has some limitations, as not all words detected are of the correct type. For example, in the sentence \textit{In reality, this is our solution}, the algorithm does not detect \textit{our} as being a pronoun. In other cases, the descriptor word of \textit{How} is not an adjective or adverb. Words such as \textit{no, You, the}, and \textit{I} are not descriptor words. In many cases, the sentence forming {\em Consequences} is incoherent. 

Despite the handling of outlier cases, the algorithm has the following limitations. In context-specific cases, ambiguity is not handled properly. Some verbs may be interpreted as nouns, and vice versa. Furthermore, pyWSD \cite{pywsd} is sometimes unsuccessful in disambiguating the meaning of a noun, causing it to be possibly mislabeled as \textit{MISC}. Therefore, the algorithm is better suited to handle general conversation topics, rather than specialized topics, like solving computer programming exercises. The choice to use a rule-based approach to detect speech clauses rather than Machine Learning is less than ideal for a final algorithm design, and it is evident that the detected clauses are not robust. Some descriptor words are not adjectives or adverbs even if they are marked as such, and sometimes there are repeated words detected in clauses {\em Why} and {\em Consequences}. Overall, this algorithm represents a first attempt at designing a basic NLP system. Still the results generated are accurate enough to generate further hypotheses regarding speaker intentions.

\subsection{Hypotheses verification on social situations}

This experiment highlighted the using of diaLogic system to verify the hypothesis suggesting that a changing interactions between team members expressed as $\Delta IG$ result in a change of the member's emotions too, $\Delta E$. The hypothesis was verified in two situations, constant and variable team structures.   

\subsubsection{Constant team structure}

The interaction dynamic results for the first video are displayed in Figure~\ref{C-1}. For the entire video duration of the first video, the two speakers with the least $\Delta IG$ were Participant 1 and Participant 3. The participant with the most $\Delta E$ was Participant 2.
For the entire video duration of the second video, the two speakers with the least $\Delta IG$ were Participant 1 and Participant 2. The participant with the most $\Delta E$ was Participant 3.

\subsubsection{Variable team structure}

\begin{figure*}[t]
	\centering
	\begin{subfigure}[t]{0.4\textwidth}
		\centering
		\includegraphics[width=\textwidth]{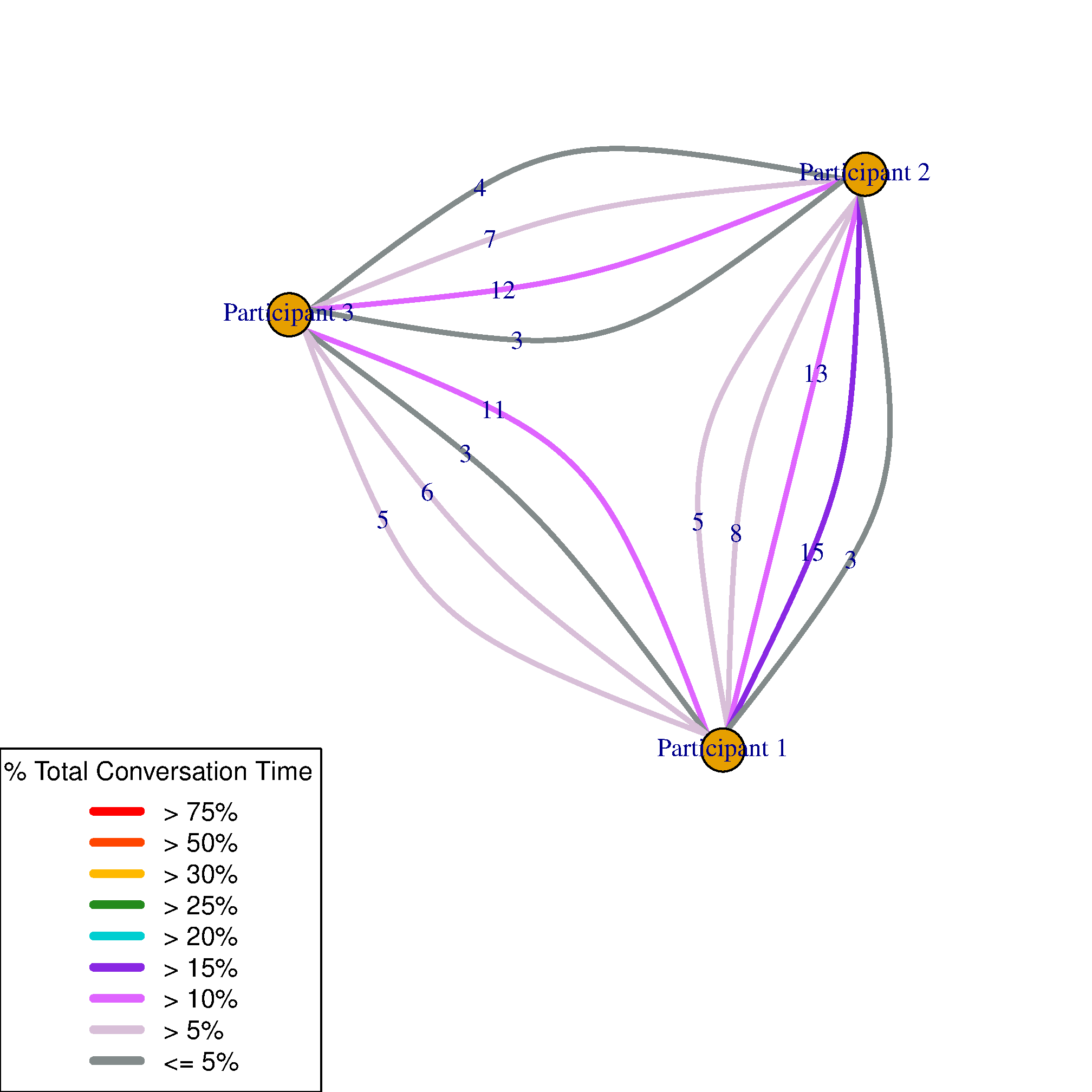}
		\caption{$T_1$ Participant Interactions 0-2 Minutes}
		\label{V-1b-1}
	\end{subfigure}
	\hfill
	\begin{subfigure}[t]{0.4\textwidth}
		\centering
		\includegraphics[width=\textwidth]{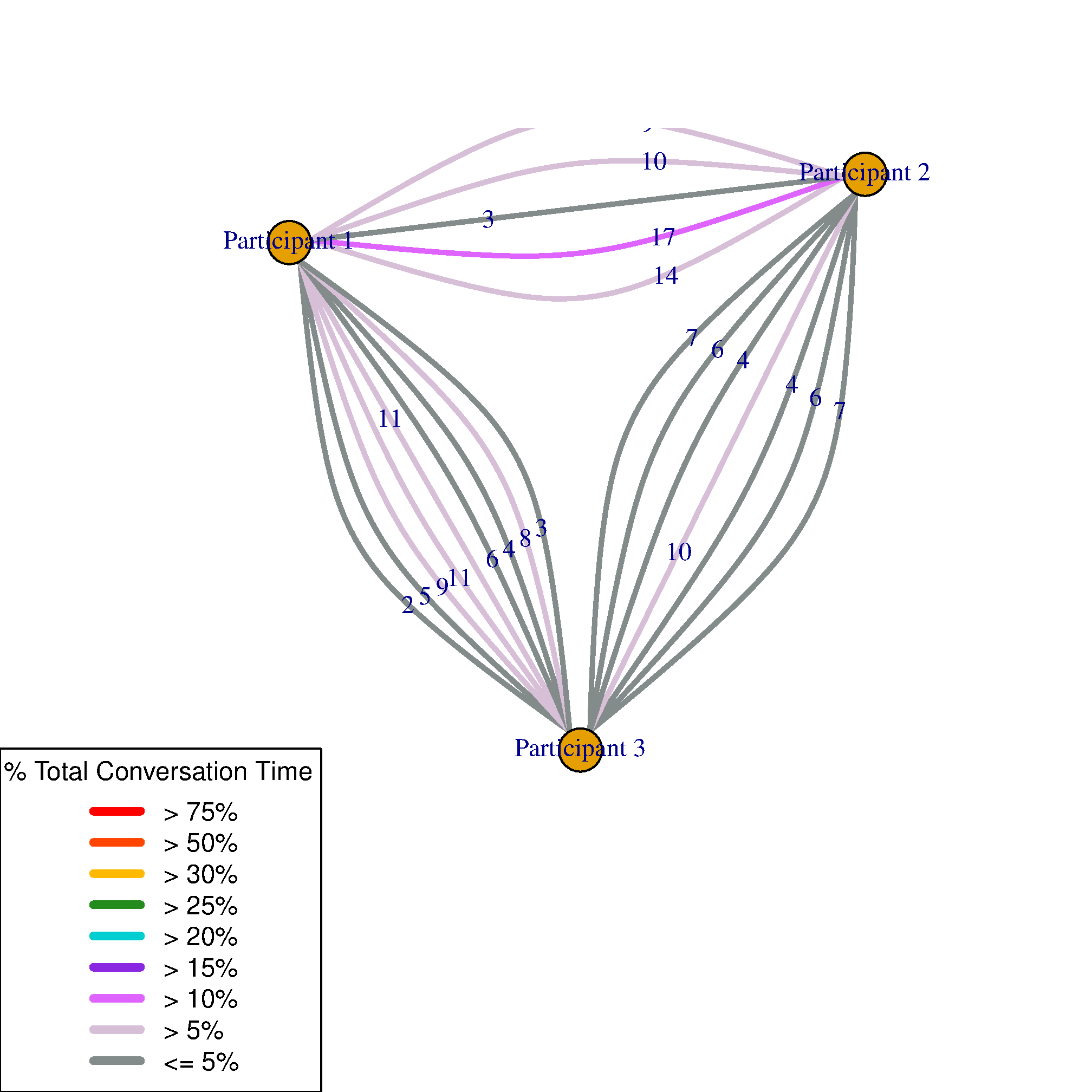}
		\caption{$T_2$ Participant Interactions 2-4 Minutes}
		\label{V-1b-2}
	\end{subfigure}
	\newline
	\begin{subfigure}[t]{0.4\textwidth}
		\centering
		\includegraphics[width=\textwidth]{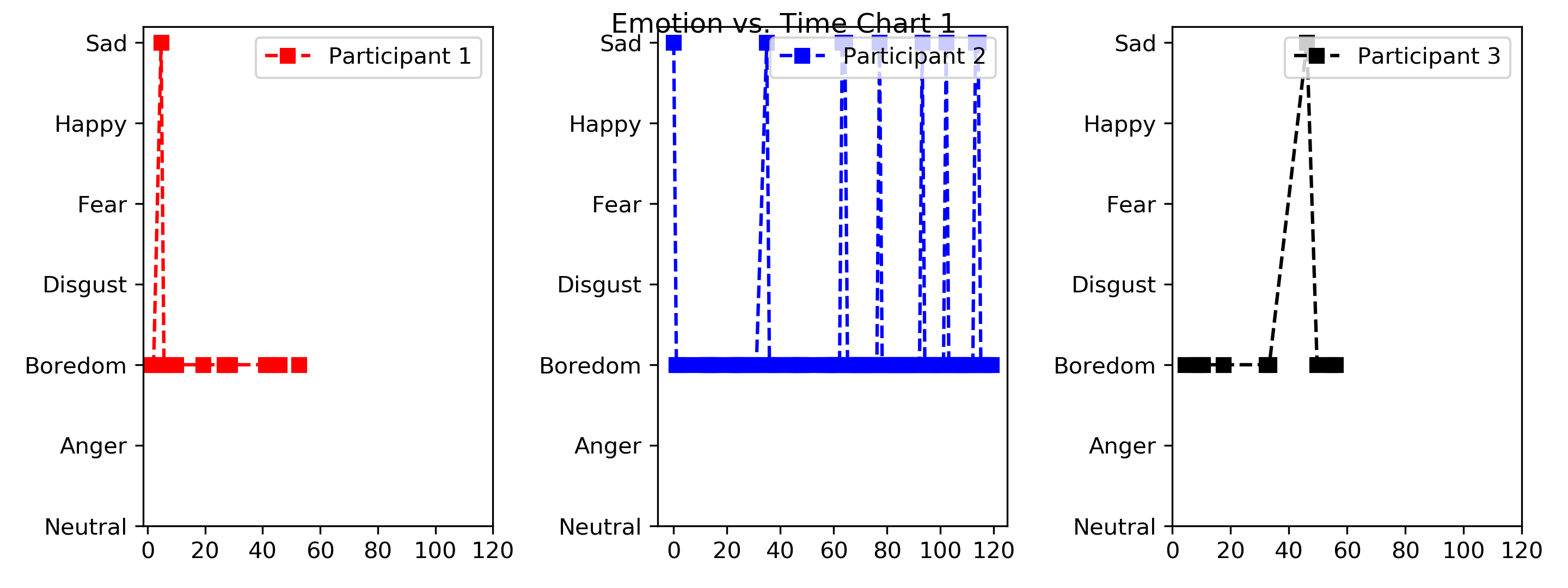}
		\caption{$T_1$ Participant Emotions 0-2 Minutes}
		\label{V-1a-1}
	\end{subfigure}
	\hfill
	\begin{subfigure}[t]{0.4\textwidth}
		\centering
		\includegraphics[width=\textwidth]{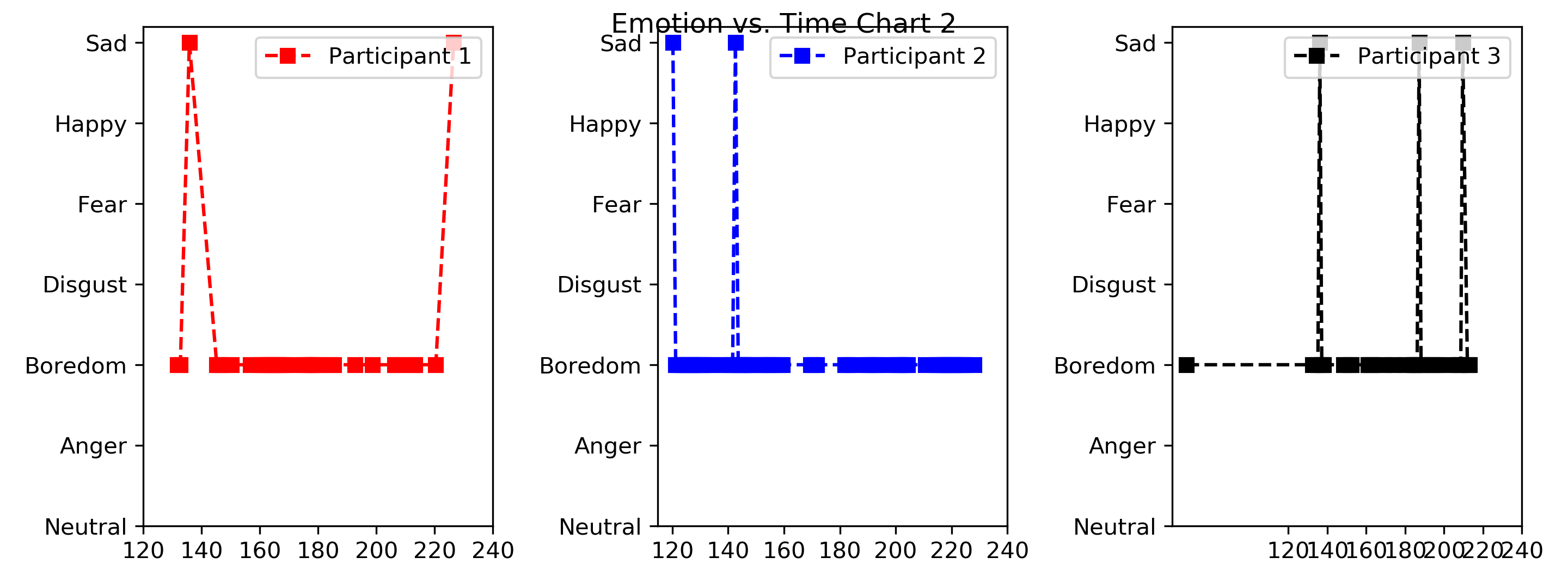}
		\caption{$T_2$ Participant Emotions 2-4 Minutes}
		\label{V-1a-2}
	\end{subfigure}
	\caption{Variable interaction dynamic}
	\label{V-1}
\end{figure*}


The interaction dynamic results for the first video with a variable dynamic are displayed in Figure~\ref{V-1}. The dynamic within these videos shifts from $T_1$ (0-2 minutes) to $T_2$ (2-4 minutes). The two speakers with the least $\Delta IG$ within $T_1$ were Participant~1 and Participant~3. The two speakers with the least $\Delta IG$ within $T_2$ were Participant~1 and Participant~2. The participant with the most $\Delta E$ within $T_1$ was Participant~2 and the participant with the most $\Delta E$ within $T_2$ was Participant~3. 
The dynamic within the second video shifts from $T_1$ (0-2 minutes) to $T_2$ (2-4 minutes). The two speakers with the least $\Delta IG$ within $T_1$ were Participant~1 and Participant~2. The two speakers with the least $\Delta IG$ within $T_2$ were Participant~2 and Participant~3. The participant with the most $\Delta E$ within $T_1$ was Participant~3 and the participant with the most $\Delta E$ within $T_2$ was Participant~2. 

In conclusion, the considered hypothesis held for the majority of the cases in the four videos, but it was not verified for the second time interval of the variable team structure situation. 

\section{Conclusions}

This paper presents diaLogic, a Human-In-A-Loop system, to model team behavior during open-ended problem solving. It performs automated multi-faceted data acquisition that can be then used to extract hypotheses about team behavior. Cognitive, emotional, and social characteristics are identified based on speech within recorded videos of social experiments. The core algorithm is a speaker diarization algorithm, from which all subsequent data are computed, like speech emotion recognition, speaker interaction, speech-to-text, and speech clause information. A rule-based algorithm identifies the types of the clauses in responses, e.g., {\em What}, {\em Who}, {\em For who}, {\em When}, {\em How}, {\em Where}, {\em Why}, and {\em Consequences} clauses. Experiments show that data acquisition accuracy is enough to support qualitative interpretation of team behavior. diaLogic system can be utilized in broad range of applications from education to team research and psychology. It offers higher accuracy when processing normal conversations, rather than conversations within a specialized context. 


Future work
will focus to expand the amount of data and hypotheses drawn from the core algorithms.
For future data classification, sarcasm, irony, enthusiasm, and confidence will be detected. 
To build these classifiers, new voice databases need to be created. This process requires a few considerations. First, the databases must contain clean audio. Second, the databases must feature accurate representations of speech properties. At the moment, possible audio sources for data are audiobook recordings. It remains to be studied whether proper content exists to create such a voice database.

\end{document}